\documentclass[]{interact}

\usepackage{epstopdf}

\usepackage{natbib}
\bibpunct[, ]{(}{)}{;}{a}{}{,}

\usepackage{tabu,hhline,multirow}
\usepackage[section]{placeins}
\usepackage{graphicx}
\usepackage{mathtools}
\usepackage{caption}
\usepackage{subcaption}
\usepackage{xcolor}

\theoremstyle{plain}

\theoremstyle{definition}

\theoremstyle{remark}

\newcommand{\ba}{\begin{align}}
\newcommand{\ea}{\end{align}}

\graphicspath{{figures/}}

\begin{document}


\title{Numerical Study of Flow Structure and Pedestrian Level Wind Comfort Inside Urban Street Canyons}


\author{
\name{P.~P. Pancholy\textsuperscript{a}, K. Clemens\textsuperscript{b}, P. Geoghegan\textsuperscript{c}, M. Jermy\textsuperscript{d}, M. Moyers-Gonzalez\textsuperscript{e} and P.~L. Wilson\textsuperscript{e,f}\thanks{CONTACT P.L. Wilson. Email: phillip.wilson@canterbury.ac.nz}}
\affil{\textsuperscript{a}Quake Centre, Department of Civil and Natural Resources Engineering, University of Canterbury, Christchurch 4800, New Zealand;
	\textsuperscript{b}Private Consultant; \textsuperscript{c}Department of Biomedical Engineering, Aston University, Birmingham, B4 7ET, UK; \textsuperscript{d}Department of Mechanical Engineering, University of Canterbury, Christchurch 4800, New Zealand; \textsuperscript{e}School of Mathematics \& Statistics, University of Canterbury, Christchurch 4800, New Zealand; \textsuperscript{f}Te P\={u}naha Matatini, New Zealand.}
}

\maketitle

\begin{abstract}
	
 In this work we numerically investigate the flow conditions inside uniform and non-uniform street canyons well within the atmospheric boundary layer. The numerical simulations use the steady RANS method with the near-wall modelling approach to simulate wall roughness at the boundary. With the aim of investigating both flow structure in broad terms, and pedestrian comfort in the street canyon between parallel buildings, we test different canyon configurations with varied street width, building width and building height. Turbulent conditions are broadly expected to hold within the physically-realistic range of Reynolds number of order $10^6$ considered here, where we take the building height to be a characteristic length scale, and the free stream velocity as the characteristic velocity. In addition to discussing the features of the canyon and wake flow, we investigate the effects of canyon geometry on pedestrian comfort by using the Extended Land Beaufort Scale for this purpose. We present and compare pedestrian comfort ``maps'' for each of our geometries.   
\end{abstract}


%
%

\section{Introduction}

Urban street canyons are places in cities or towns where a street is flanked on both sides by a continuous or almost continuous line of buildings. It is known that these street canyons affect local variations in global environmental and meteorological conditions such as wind speed, wind direction, air pollution, rain direction, and heat radiation \citep{key-3}. The goal of the present work is to study how the geometrical features of the canyon and the buildings which define it affect the wind flow structure inside it in order to study not only the broad features of wind canyon flow and their dependence on canyon geometry, but mainly to study the effects of canyon geometry on pedestrian comfort. Our canyons consist of two large buildings of the same spanwise length and streamwise width, representing either identical parallel low buildings or, more likely, quasi-uniform terraced buildings on either side of a street.

\begin{figure}
	\vspace*{-6mm}
	\captionsetup[subfigure]{aboveskip=-2pt,belowskip=-2pt}
	\centering
	\begin{subfigure}[c]{\textwidth}
		\centering
		\includegraphics[width=\linewidth]{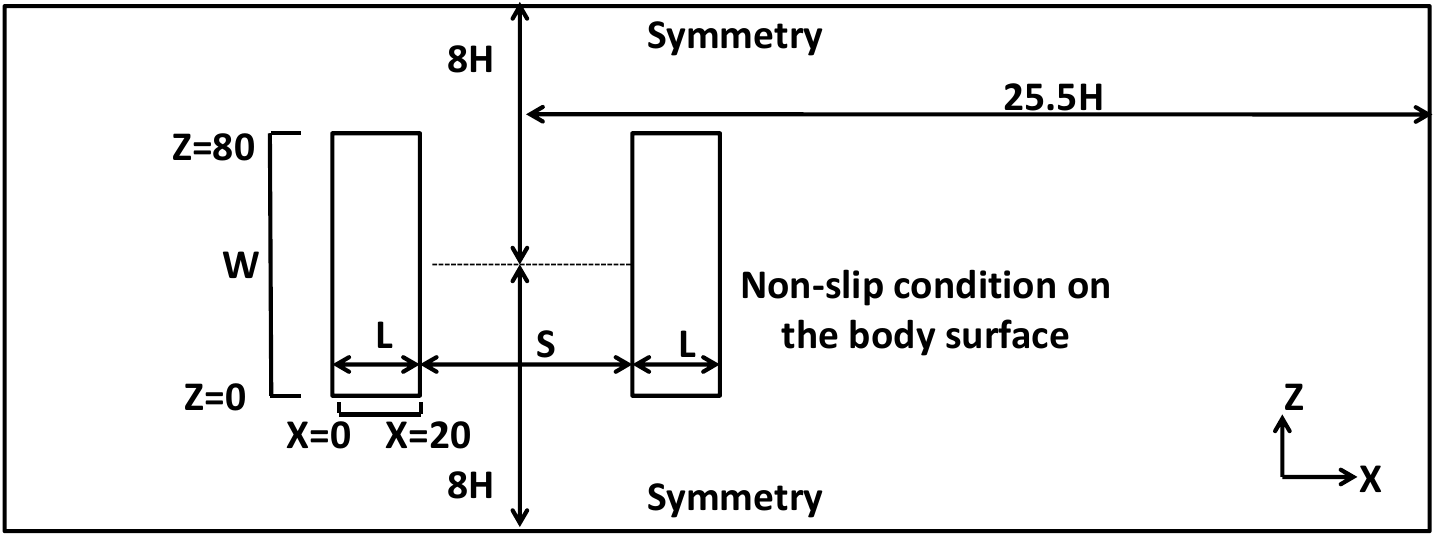}
		\caption{}
		\label{fig:sub4.1.1}
	\end{subfigure}\hfill
	\begin{subfigure}[c]{\textwidth}
		\centering
		\includegraphics[width=\linewidth]{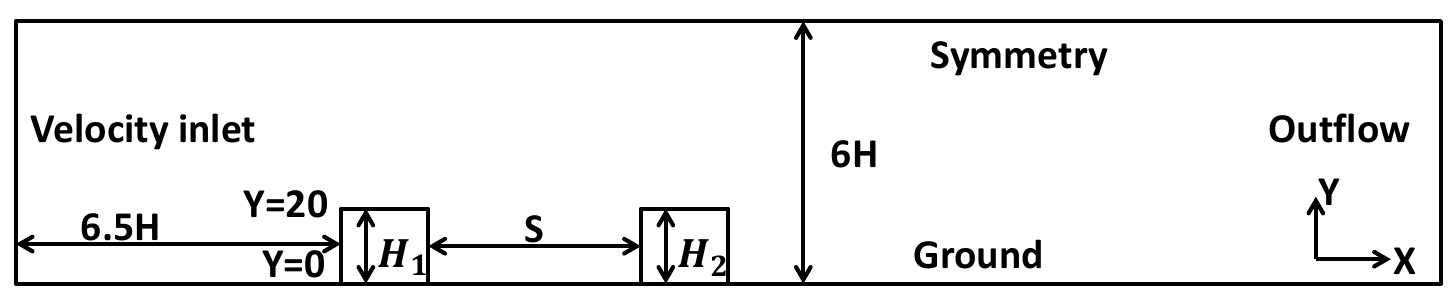}
		\caption{}
		\label{fig:sub4.1.2}\hfill
	\end{subfigure}
	\vspace*{-3mm}
	\captionsetup[figure]{font=small,skip=0pt}
	\caption{Computational domain and boundary conditions for the uniform street canyon (A) plan view (B) side view. }
	\label{Fig:4.1}
\end{figure}

The detailed canyon geometry is explained in Section \ref{sec:CFD simulations: computational model and parameters for uniform street canyon study cases}, but for the purpose of the introduction we require the geometrical parameters of the street width ($S$), the height of the upstream building ($H_1$) and that of the downstream building ($H_2$), as shown later in Figure~\ref{Fig:4.1}. A street canyon is said to be \emph{uniform} if the adjacent building heights are equal, that is, $H_{1}=H_{2}$. If the building heights are not equal, the canyon is said to be \emph{non-uniform}, and either \emph{step-up} if $H_{1}< H_{2}$ or \emph{step-down} if $H_{1}> H_{2}$ . The dimensions of a street canyon are broadly expressed by its aspect ratios $S/H$ and $S/W$, where $H$ is the height of the tallest building and $W$ the maximum width of the buildings.

A further subclassification can be defined as follows:
\begin{itemize}
	\item Deep canyon if $S/H\approx 1/2$;
	\item Regular canyon if $S/H\approx 1$;
	\item Avenue canyon if $S/H\approx 2$;
	\item Long canyon if $S/W\approx 1/7$;
	\item Medium canyon if $S/W\approx 1/5$;
	\item Short canyon if $S/W\approx 1/3$.
\end{itemize}

\subsection{Flow structure around tandem obstacles}

Flow past obstacles has been widely studied both experimentally and numerically. For the case of a single symmetric obstacle with oncoming wind being normal to the span-wise direction, the flow separates upstream of the windward face or faces. The separated shear layers may or may not reattach at the top of the obstacle. Reattachment or continued separation of these separated shear layers is of the utmost importance as it alters the wake flow periodicity which changes the pressure and velocity field near the obstacle. In the case of a geometrically simple obstacle, such as the rectangular prisms considered here, the reattachment depends on the boundary layers thickness $\delta$ and the length to height ratio ($L/H$) of the obstacle. If $L/H\ll\delta$, the shear layers do not reattach and an extended recirculation region is present, see \citet{key-11}, \citet{key-11}, and \citet{key-12}.

\begin{figure}
	\vspace*{-1mm}
	\centering
	\includegraphics[width=\textwidth]{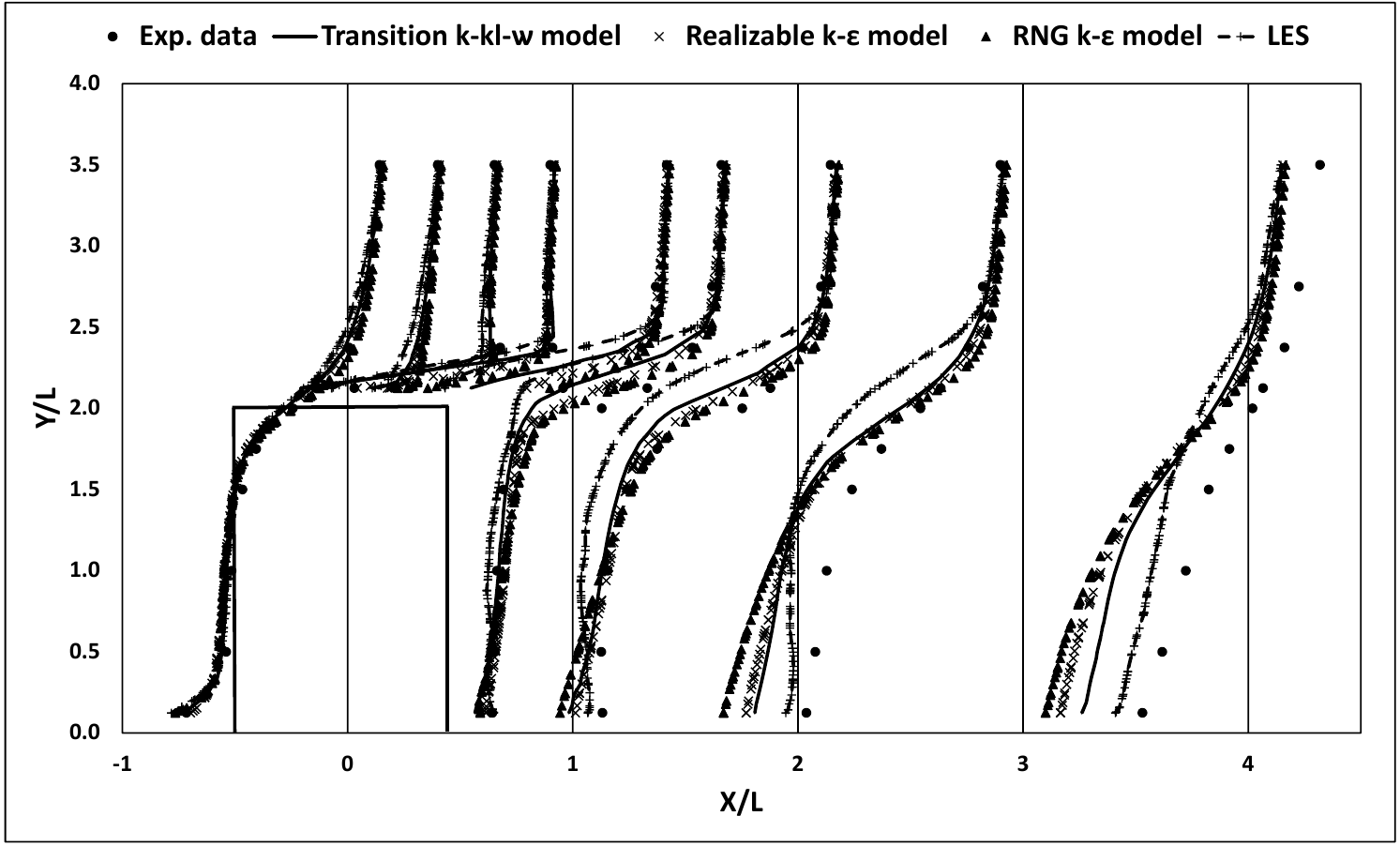}
	\vspace*{-3mm}
	\captionsetup[figure]{font=small,skip=0pt}
	\caption{Vertical U profile (at the center plane) CFD steady-RANS calculation results comparison with the experimental data points \citep{key-47} and LES results. Here $L$ represents dimension of the building in the direction of the flow.}
	\label{Fig:3.10}
\end{figure}


The flow features become more complex when instead of a single surface-mounted obstacle there is an array of obstacles \citep{key-37}. For example, for two obstacles in tandem, one in the wake of the other, the features of the flow field of the downstream obstacle which are dependent on its own wake structure are in addition dependent on the periodic flow characteristics of the wake of the upstream obstacle. This phenomenon is called \emph{buffeting} \citep{key-18}.

\citet{key-27} studied the buffeting of surface-mounted tandem cubic obstacles, identifying three broad regimes, namely: ($1$) a bistable regime similar to the unstable reattachment regime of \cite{key-37}; ($2$) a ``lock-in'' regime in which the upstream shear layer impinges on the leading face (near the leading edge) of the downstream obstacle and a strong vortex rolls in the gap between the obstacles; and ($3$) a quasi-isolated regime similar to the unstable synchronized regime of \cite{key-37}. 

The studies of \citet{key-27} and \citet{key-37}  are performed at moderate $Re$ somewhat lower than the values considered here. Moreover, these investigations were based on the reattachment of the shear layers either from the sides or the top leading edges of the upstream obstacle, but not both. Experimental and numerical simulations to analyse the flow structure for tandem obstacles at very high $Re$ (in the range of $10^{6} - 10^{7}$) inside a thick boundary layer with reattachment of all three separated shear layers have yet to be reported in the literature. This type of flow study is important because urban wind flow around buildings typically occurs at the Reynolds numbers we consider herein, and moreover do so within the thick atmospheric boundary layer. The present parametric study aims to fill this gap by studying the the $3D$ flow structure around and in-between two medium-rise parallel buildings of simple cross-section forming a street canyon. In addition to the overall flow structure, of interest from a purely fluid dynamical point of view, we are also interested in the effects of the complex flow patterns on the comfort of pedestrians in the street, as we explain the next subsection.

\begin{figure}
	\vspace*{-5mm}
	\centering
	\includegraphics[width=\textwidth]{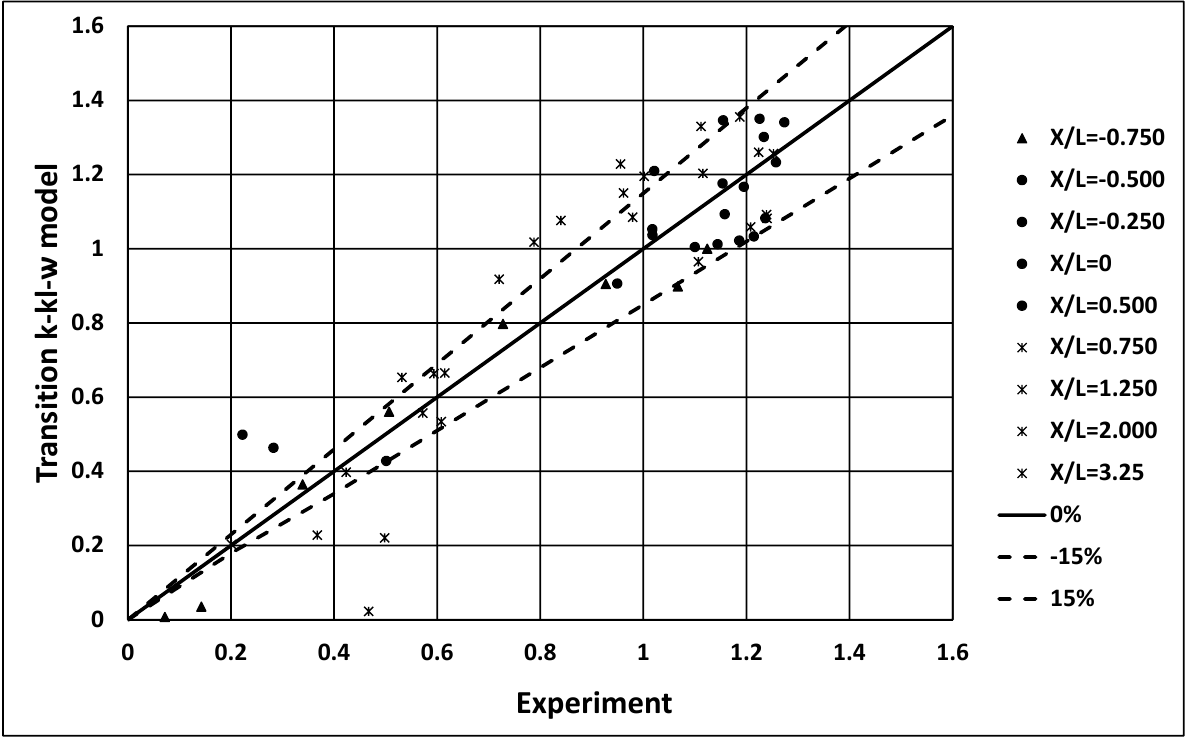}
	\vspace*{-3mm}
	\captionsetup[figure]{font=small,skip=0pt}
	\caption{Comparison of wind speed increase ratio obtained using $k$-$kl$-$\omega$ model with experimental data points \citep{key-47} in a plane near the ground. $X/L=-0.75$ is upstream of the building, whereas $X/L=-0.500$, $-0.250$, $0$, and $0.5$ are on the sides of the building, and $X/L=0.750$, $1.250$, $2.000$, and $3.25$ are downstream.}
	\label{Fig:3.13}
\end{figure}

\subsection{Pedestrian comfort inside uniform and non-uniform street canyon}

Urban areas can be characterized as a set of street canyons. Wind comfort and wind safety for pedestrians are important requirements for the development of urban areas. In the early 1800s Sir Francis Beaufort devised an empirical table which relates wind speed measured at 10 meters above the ground to the observed conditions. More than a century and a half later, \citet{key-24} extended the Beaufort scale to create the Extended Land Beaufort Scale, or ELBS, which related the average wind speed at a representative pedestrian height of 1.75m with a description of pedestrian comfort. For example, at ELBS number 3 (a gentle breeze or wind speed of 2.4--3.8m/s) individuals will experience disturbed hair and flapping clothes, with newspapers becoming difficult to read. In the present work, an ELBS number of 3 and above will be considered as unpleasant for pedestrians. 


According to \citet{key-4}, fundamental studies for pedestrian level wind comfort such as the present work are typically conducted for simple, generic building configurations to obtain insight into the flow behaviour, study the influence of different building dimensions and street widths, provide input for knowledge-based expert systems (KBES), and for model validation. Experimental studies of this kind have been conducted by \citet{key-22} and \citet{key-45}, who carried out wind tunnel measurements along the street centre line in various two-building configurations.  Both studies focused on the mean wind speed in the street between parallel rectangular buildings of equal height. More detailed wind-tunnel measurements were given by \citet{key-39}, who reported contours of mean wind speed and turbulence measurements at pedestrian level in street canyons between two buildings of equal height for both parallel and perpendicular wind direction, with respect to the upstream face of the buildings. Numerical studies for canyons defined by two parallel buildings were conducted by \citet{key-8} and \citet{key-2}. A very detailed numerical assessment of the influence of varying a wide range of street widths was first conducted by \citet{key-4} for parallel wind direction and with buildings of equal height.

These studies on wind speed conditions in a street canyon were mainly focused on pedestrian-level winds at discrete points and a rather limited range of street widths. To the best of our knowledge, we are the first to present a wind-comfort assessment for both uniform and non-uniform street canyons across a wide range of wind and geometric conditions and to provide the mean wind speed at the pedestrian level across the entire street for a perpendicular wind direction in order to assess pedestrian wind comfort inside street canyons.

In Section \ref{sec:CFD simulations: computational model and parameters for uniform street canyon study cases}, we  discuss the model, boundary conditions and numerical validation. In Section \ref{sec:Results and discussion for uniform street canyons} and  \ref{sec:Results and discussion for non-uniform street canyons} we analyse the flow structure and pedestrian comfort inside uniform and non-uniform street canyons respectively. Finally, in Section~\ref{sec:Conclusion} we present the conclusion of this study.

\begin{figure}[t]
	\vspace*{-2mm}
	\captionsetup[subfigure]{aboveskip=-2pt,belowskip=-2pt}
	\centering
	\begin{subfigure}[c]{0.48\textwidth}
		\centering
		\includegraphics[width=\linewidth]{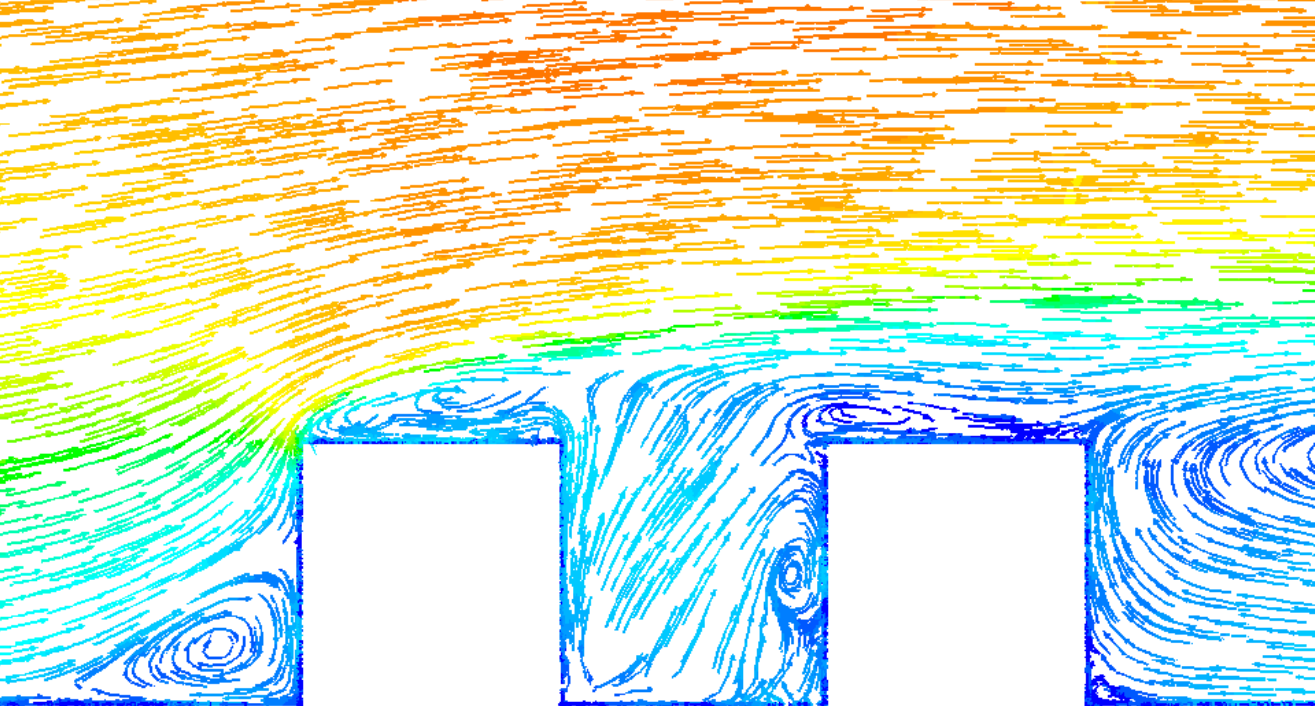}
		\caption{$S/H = 1$}
		\label{fig:sub4.6.4}
	\end{subfigure}\hfill
	\begin{subfigure}[c]{0.48\textwidth}
		\centering
		\includegraphics[width=1\linewidth]{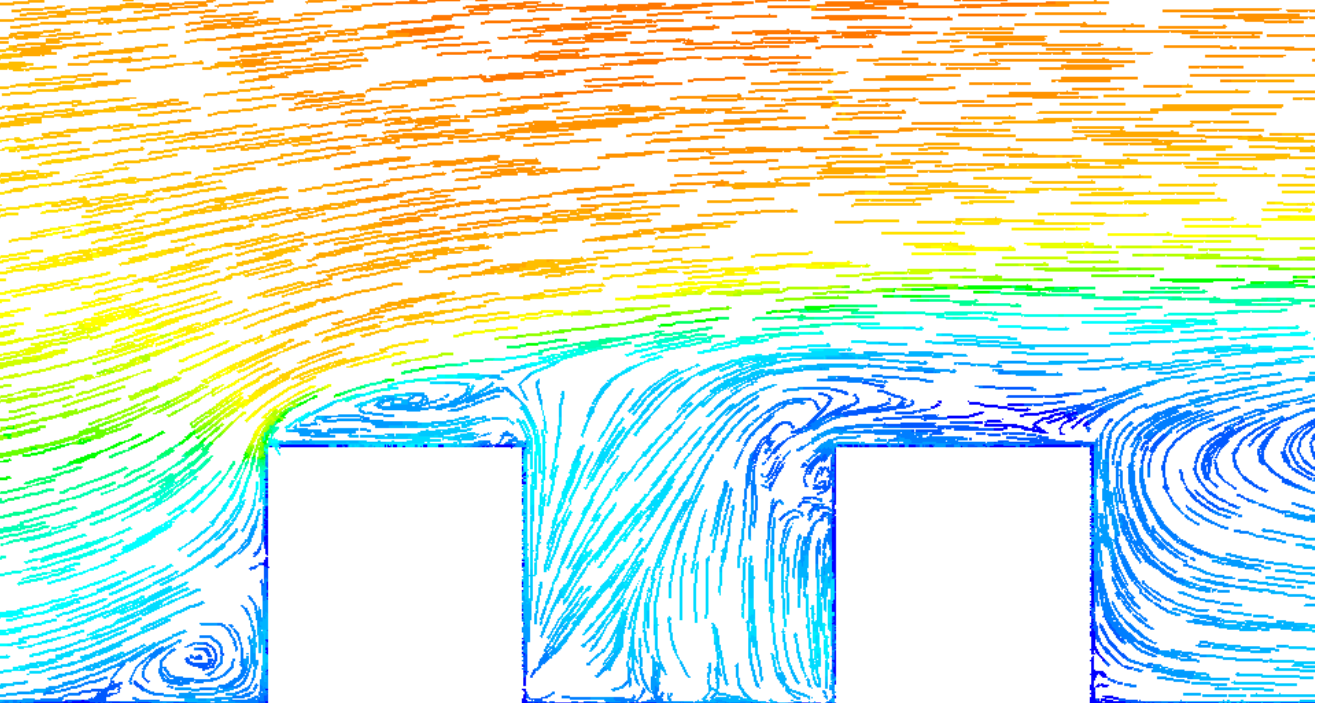}
		\caption{$S/H = 1.2$}
		\label{fig:sub4.6.5}
	\end{subfigure}\hfill
	\begin{subfigure}[c]{0.48\textwidth}
		\centering
		\includegraphics[width=1\linewidth]{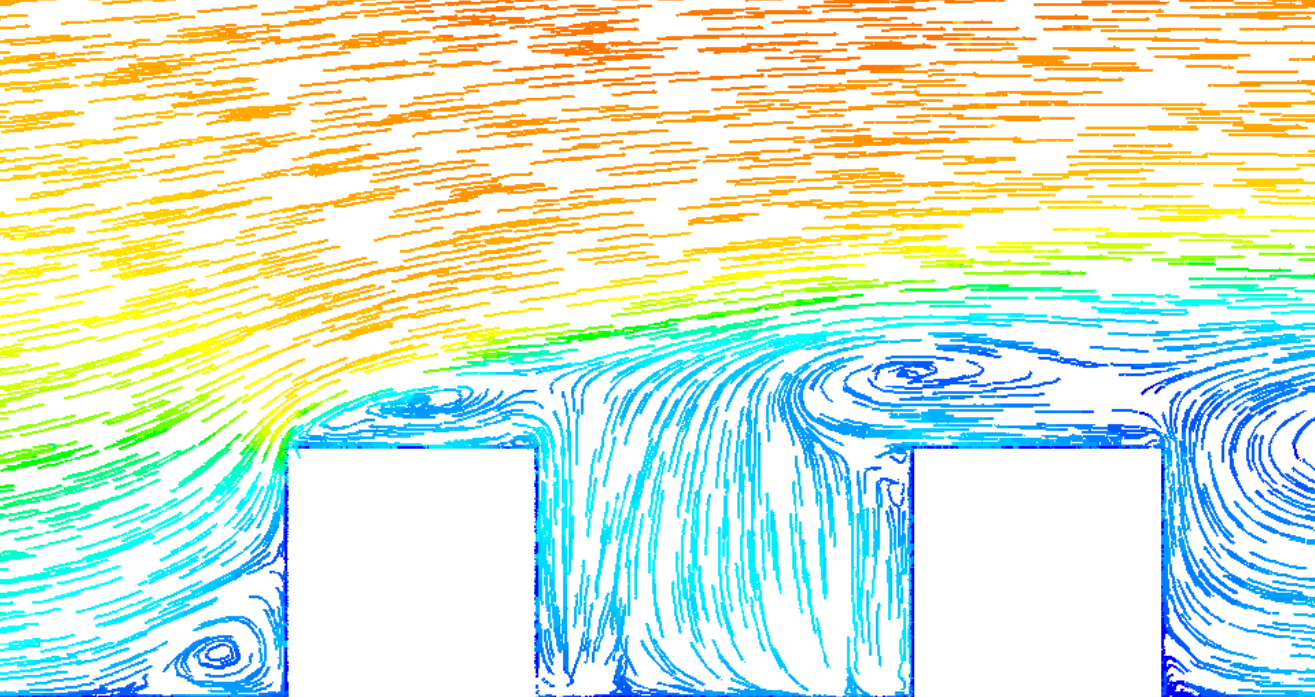}
		\caption{$S/H = 1.5$}
		\label{fig:sub4.6.6}
	\end{subfigure}\hfill
	\begin{subfigure}[c]{.48\textwidth}
		\centering
		\includegraphics[width=1\linewidth]{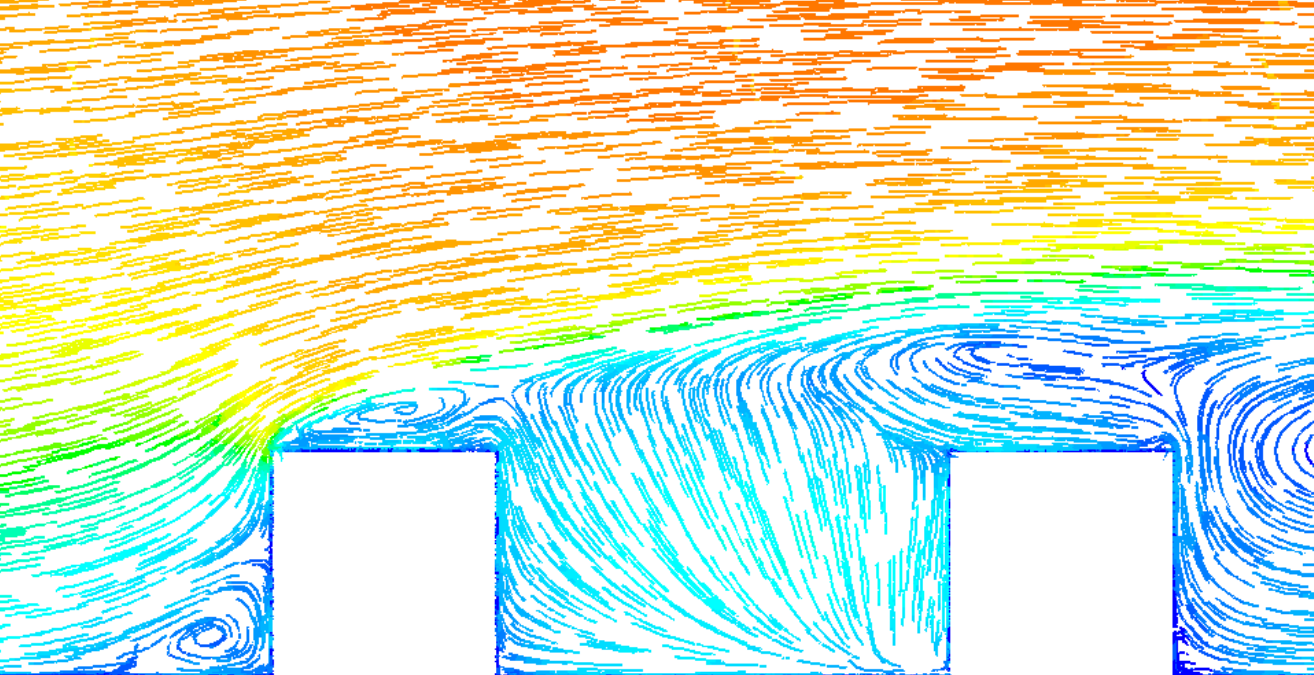}
		\caption{$S/H = 2$}
		\label{fig:sub4.6.7}
	\end{subfigure}
	\vspace*{-3mm}
	\captionsetup[figure]{font=small,skip=0pt}
	\caption{Velocity magnitude pathlines in the $XY$-plane at the centre plane ($Z = 40$) for the uniform street canyon with different street aspect ratios and $W/H = 4$.}
	\label{Fig:4.6}
\end{figure}

%
%

\section{CFD simulations: computational model and parameters for uniform and non-uniform street canyon study cases}
\label{sec:CFD simulations: computational model and parameters for uniform street canyon study cases}
\subsection{Geometrical considerations}
\label{subsec:Model description for street canyon}

In Figure~\ref{Fig:4.1} we present the layout of a model street canyon. For the rest of the paper we fix the length of the buildings to be $L=20$m. We will vary the rest of the geometrical parameters as follows.
\begin{itemize}
	\item Uniform street canyon with variable street width. The buildings have the following dimensions, $W \times H \times L = 80$m$\times 20$m$\times 20$m. We will consider the effect of street width has on the flow field, hence we take $S = 20$, $24$, $30$, and $40$m, corresponding to medium canyons, both regular and avenue type.
	\item  Uniform street canyon with variable building width. We fix $S=H=20$m and take $W = 60$ and 100m.
	\item Non-uniform canyon. We consider $H_{1} = 16$ m and $H_{2} = 20$ m for the step-up street canyon, and $H_{1} = 20$ m and $H_{2} = 16$ m for the step-down street canyon, with street width of $S = 20$m. The width ($W$) of both buildings was kept constant at $80$m. 
\end{itemize}

 
The Reynolds number is defined as
\[ Re=\frac{\rho U_{\text{ref}} y_{\text{ref}}}{\mu},\]
where $\rho$ is the air density, $U_{\text{ref}}$ is the reference wind speed, $y_{\text{ref}}$ is a reference building height, and $\mu$ is the kinematic viscosity of the air. As outlined below, we validate our numerical approach over a range of Reynolds numbers, while the results section principally focusses on a Reynolds number of order $10^6$.

\begin{figure}
	\vspace*{-5mm}
	\captionsetup[subfigure]{aboveskip=-2pt,belowskip=-2pt}
	\centering
	\begin{subfigure}[c]{.48\textwidth}
		\centering
		\includegraphics[height=6cm]{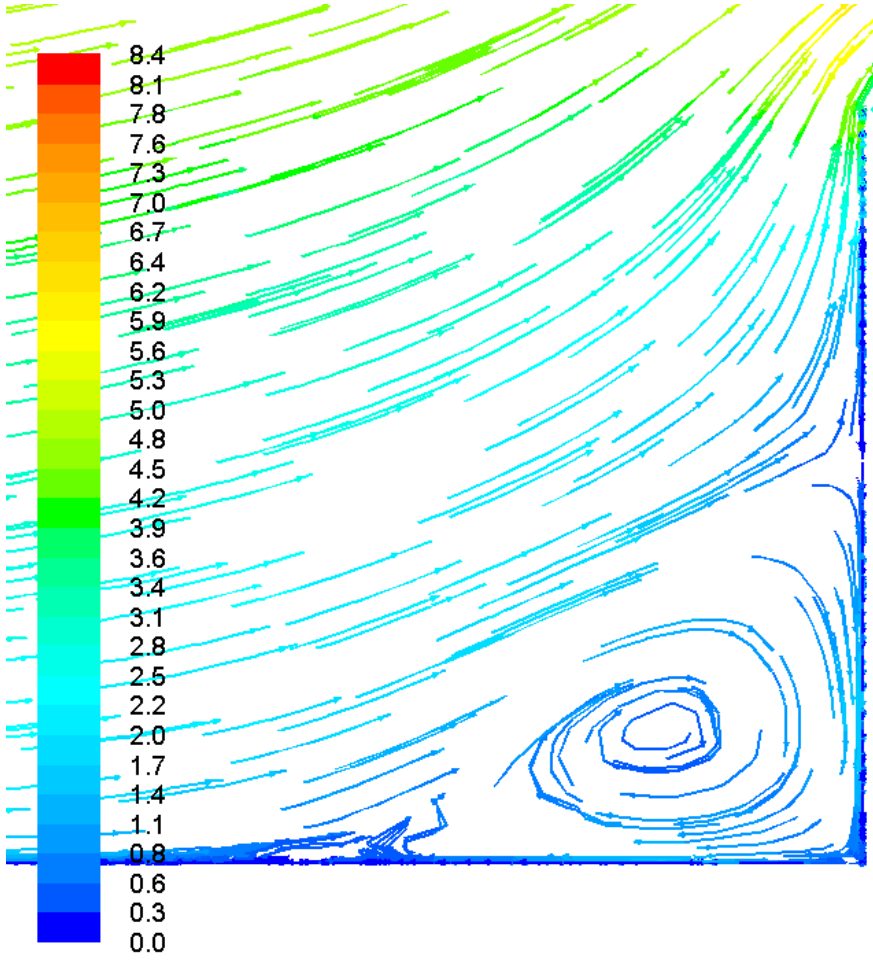}
		\caption{}
		\label{fig:sub4.7.1}
	\end{subfigure}\hfill
	\begin{subfigure}[c]{.48\textwidth}
		\centering
		\includegraphics[height=6cm]{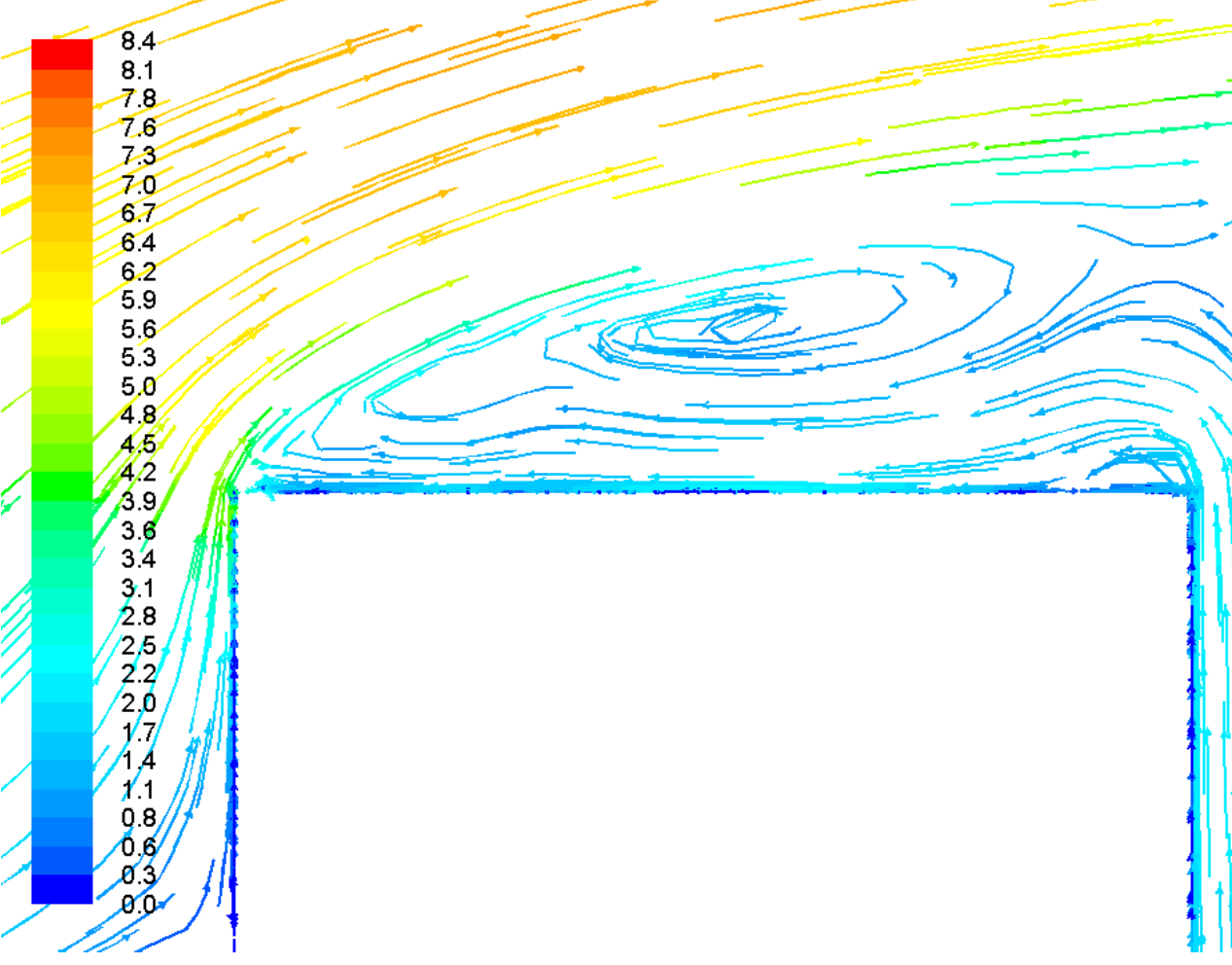}
		\caption{}
		\label{fig:sub4.7.2}
	\end{subfigure}\hfill
	\vspace*{-3mm}
	\captionsetup[figure]{font=small,skip=0pt}
	\caption{Detailed views of (A) the flow structure upstream of the upstream building, and (B) the flow separation on the roof of the upstream building; using velocity magnitude (m/s) pathlines in the $XY$-plane at the centre plane ($Z = 40$) for the street aspect ratio $S/H = 1$ and $W/H = 4$.}
	\label{Fig:4.7}
\end{figure}

\begin{figure}
	\vspace*{-5mm}
	\captionsetup[subfigure]{aboveskip=-2pt,belowskip=-2pt}
	\centering
	\begin{subfigure}[c]{.48\textwidth}
		\centering
		\includegraphics[width=\linewidth]{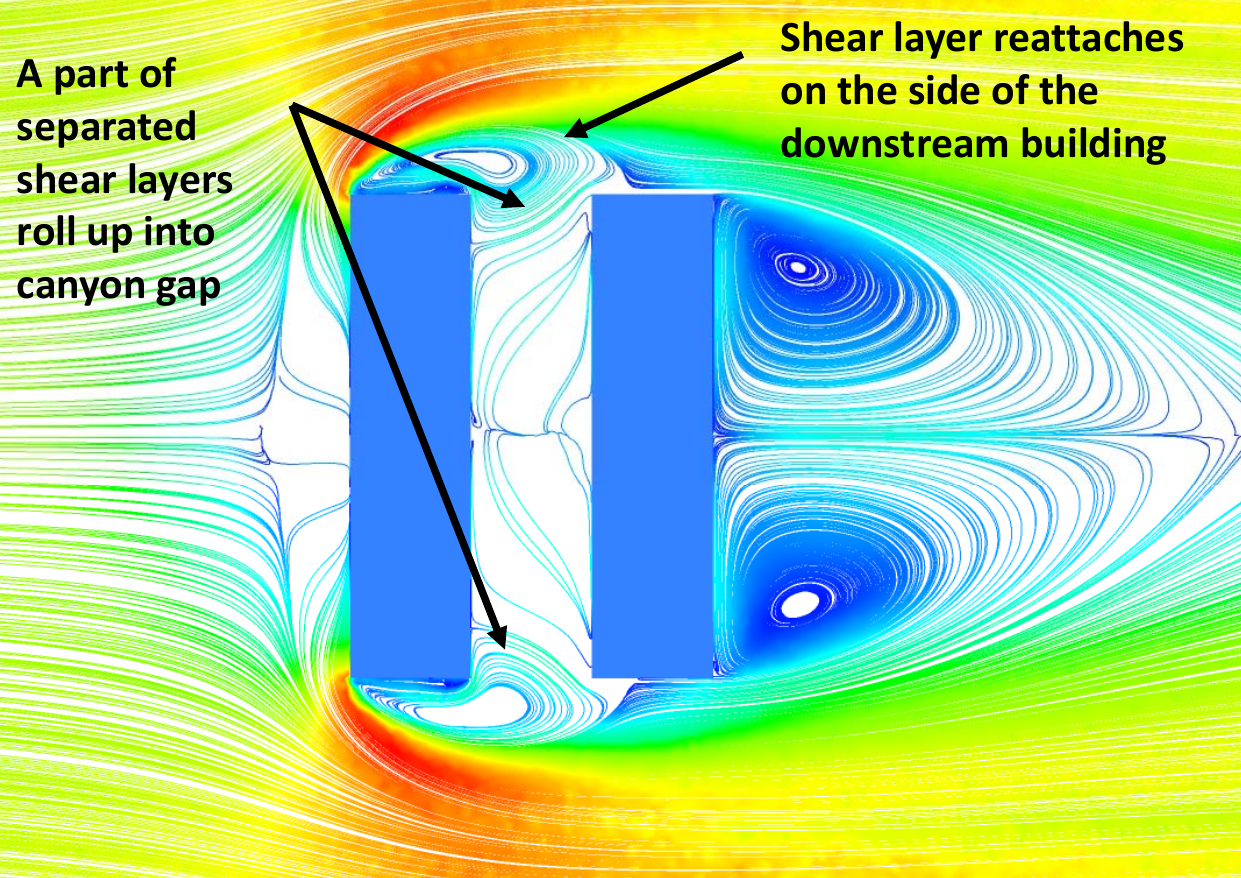}
		\caption{$S/H = 1$}
		\label{fig:sub4.8.4}
	\end{subfigure}\hfill
	\begin{subfigure}[c]{0.48\textwidth}
		\centering
		\includegraphics[width=\linewidth]{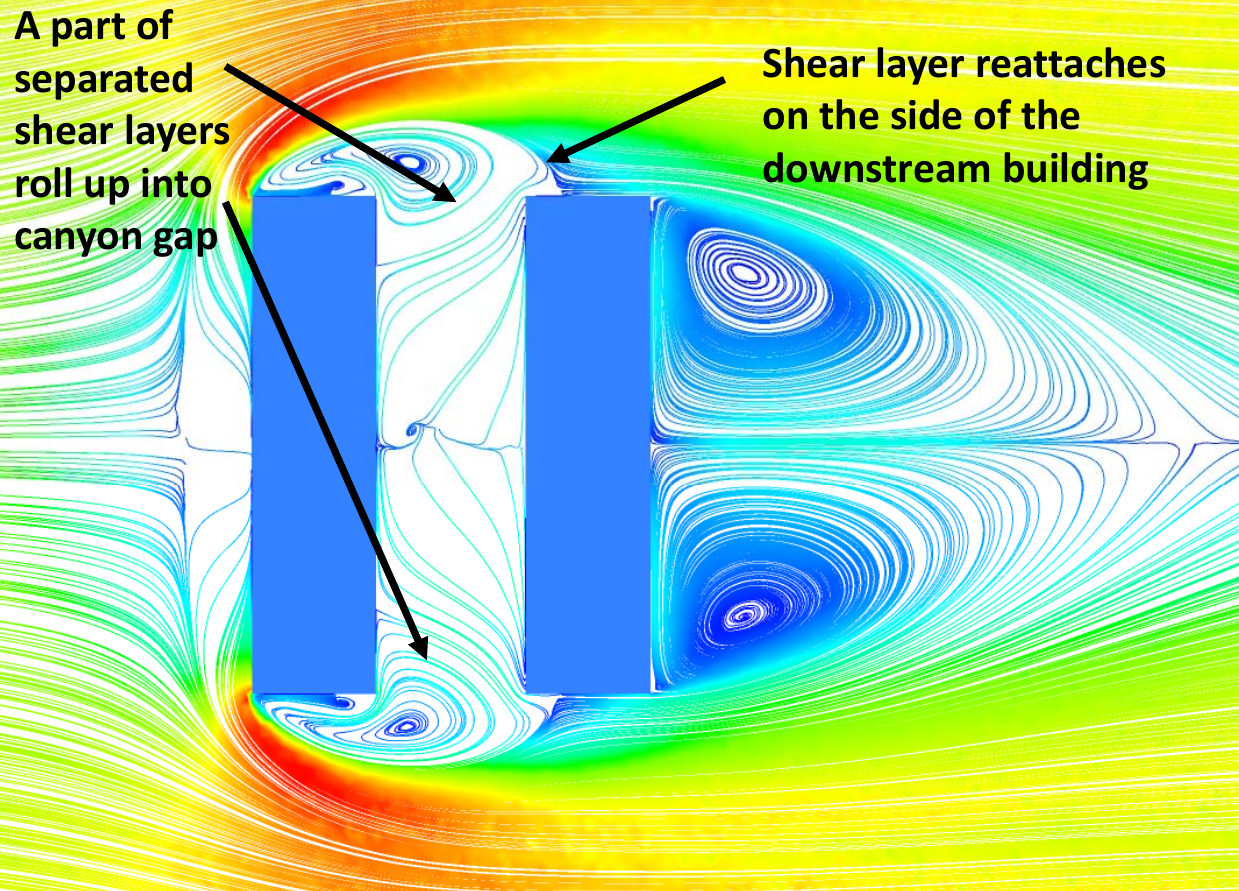}
		\caption{$S/H = 1.2$}
		\label{fig:sub4.8.5}
	\end{subfigure}\hfill
	\begin{subfigure}[c]{0.48\textwidth}
		\centering
		\includegraphics[width=\linewidth]{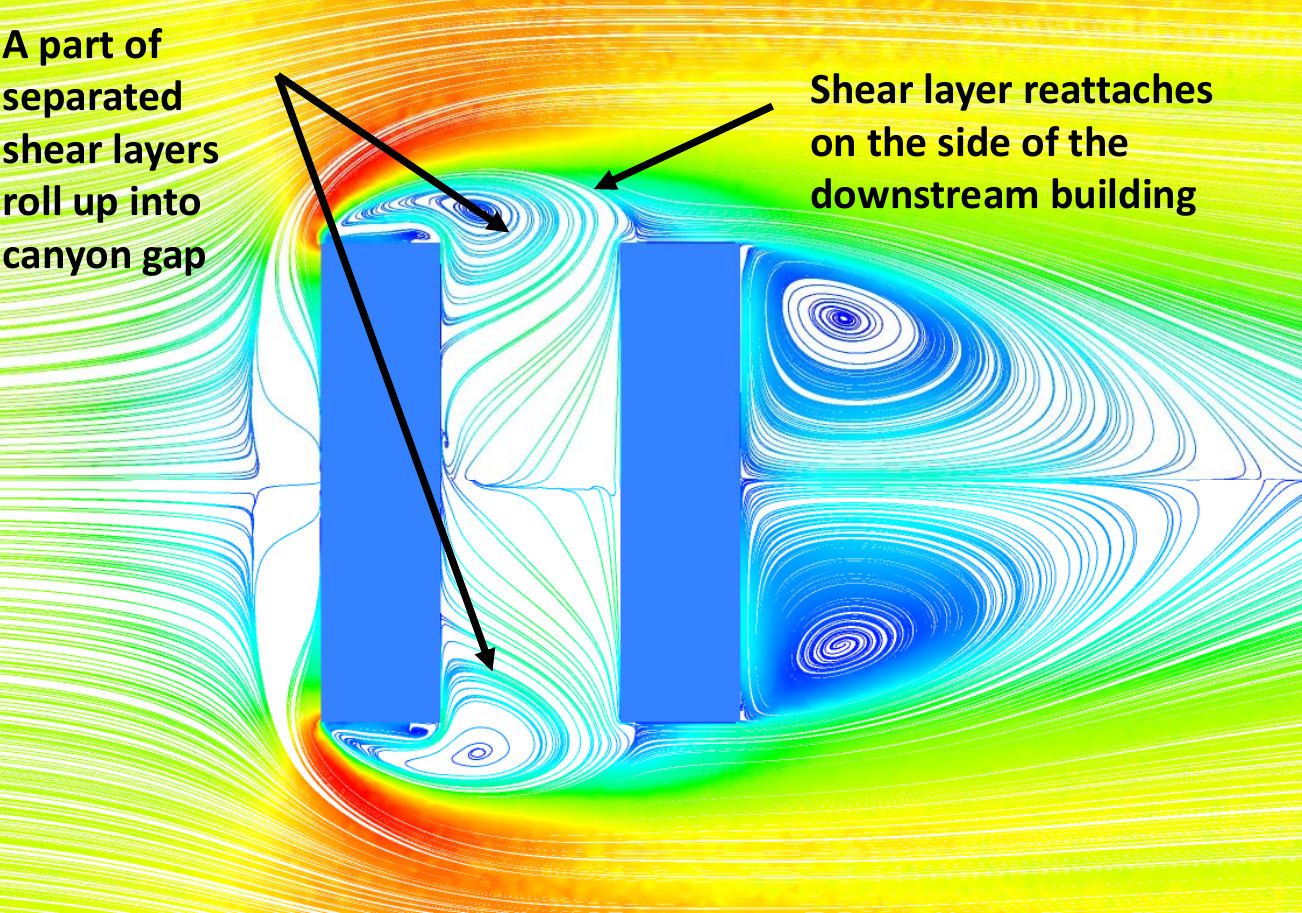}
		\caption{$S/H = 1.5$}
		\label{fig:sub4.8.6}
	\end{subfigure}\hfill
	\begin{subfigure}[c]{.48\textwidth}
		\centering
		\includegraphics[width=\linewidth]{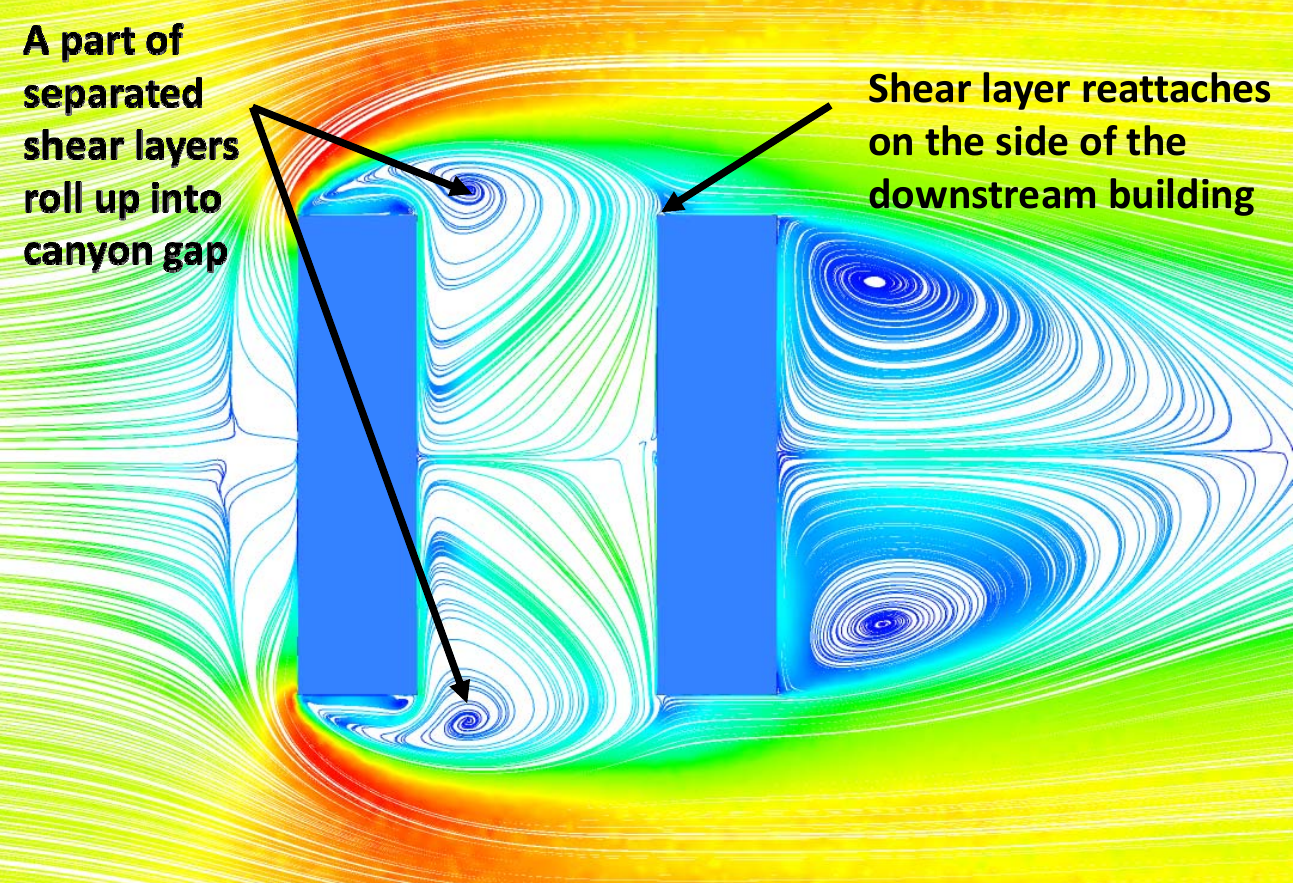}
		\caption{$S/H = 2$}
		\label{fig:sub4.8.7}
	\end{subfigure}
	\vspace*{-3mm}
	\captionsetup[figure]{font=small,skip=0pt}
	\caption{Velocity magnitude streamlines in the $XZ$-plane at $Y = 1.5$m for the uniform street canyon with different street aspect ratios and $W/H = 4$. }
	\label{Fig:4.8}
\end{figure}

\subsection{Inlet and boundary conditions and computational mesh}
\label{subsec:Boundary conditions for uniform street canyon}

The size of the computational domain was selected according to the CFD best practice guidelines given by \citet{key-17}. The domain is a box with the following dimensions: $33H$ in the $X-$direction, $6H$ in the $Y-$direction, and $16H$ in the $Z-$direction. These dimensions allow for the full development of the turbulent layer upstream of the buildings, and the development of the wake and near-building flow structures without interference from the computational domain boundaries.

\subsubsection{Inlet conditions}

The street canyons of interest are situated deep within the Atmospheric Boundary Layer (ABL), and indeed are taken to be within the logarithmic wall layer of the ABL. As such, we set the inlet velocity to be that of a logarithmic layer within a fully-developed turbulent boundary layer, such that
\begin{equation}\label{Equation:1}
U(y)=\frac{U^{*}_{\text{ABL}}}{\kappa} \ln\left(\frac{y+y_{0}}{y_{0}}\right),
\end{equation}
where $\kappa=0.42$ is the von K$\acute{a}$rm$\acute{a}$n constant, $y_{0}$ is the aerodynamic roughness length modelling upstream terrain, $C_{\mu}$ is an empirical constant, approximately equal to 0.09. Finally, $U^{*}_{\text{ABL}}$ is the ABL friction velocity, which can be calculated by a specified velocity $U_{\text{ref}}$ at reference height $y_{\text{ref}}$:
\[U^{*}_{\text{ABL}}=\frac{\kappa U_{\text{ref}}}{\ln\left(\frac{y_{\text{ref}}+y_{0}}{y_{0}}\right)}.\]
Here we take $U_{\text{ref}}= 5.9$ m/s, the free stream wind speed at the building height $y_{\text{ref}}=20$m to analyse the wind speed ranging over 2.4--3.8 m/s at pedestrian head height. Indeed, this choice of reference scales gives an ELBS number of $3$ at pedestrian height of approximately $1.75$m --- see section \ref{sec:Results and discussion for uniform street canyons}.

We also model the turbulent kinetic energy as
\begin{equation}
k(y)=\frac{3}{2}(U_{\text{avg}}I_{U})^{2},\label{kineticenergyinlet}
\end{equation}
where $U_{\text{avg}}$ is the mean wind speed at the inlet and $I_{u}$ is the longitudinal turbulence intensity, which we specify to be $21.5\%$ at the building height for all simulations. This specified turbulence intensity was obtained from AS/NZS $1170.2:2011$ \citep{key-15} at the reference building height of $20$ m.

In order to close our model we set the specific dissipation rate $\omega(y)$ and the turbulent dissipation rate $\varepsilon(y)$ as follows:
\begin{equation}\label{Equation:2}
\varepsilon(y)=\frac{U^{*3}_{\text{ABL}}}{\kappa(y+y_{0})};
\end{equation}
\begin{equation}\label{Equation:3}
\omega(y)=\frac{\varepsilon(y)}{C_{\mu}k(y)}.
\end{equation}

\begin{figure}
	\vspace*{-5mm}
	\centering
	\includegraphics[width=\textwidth]{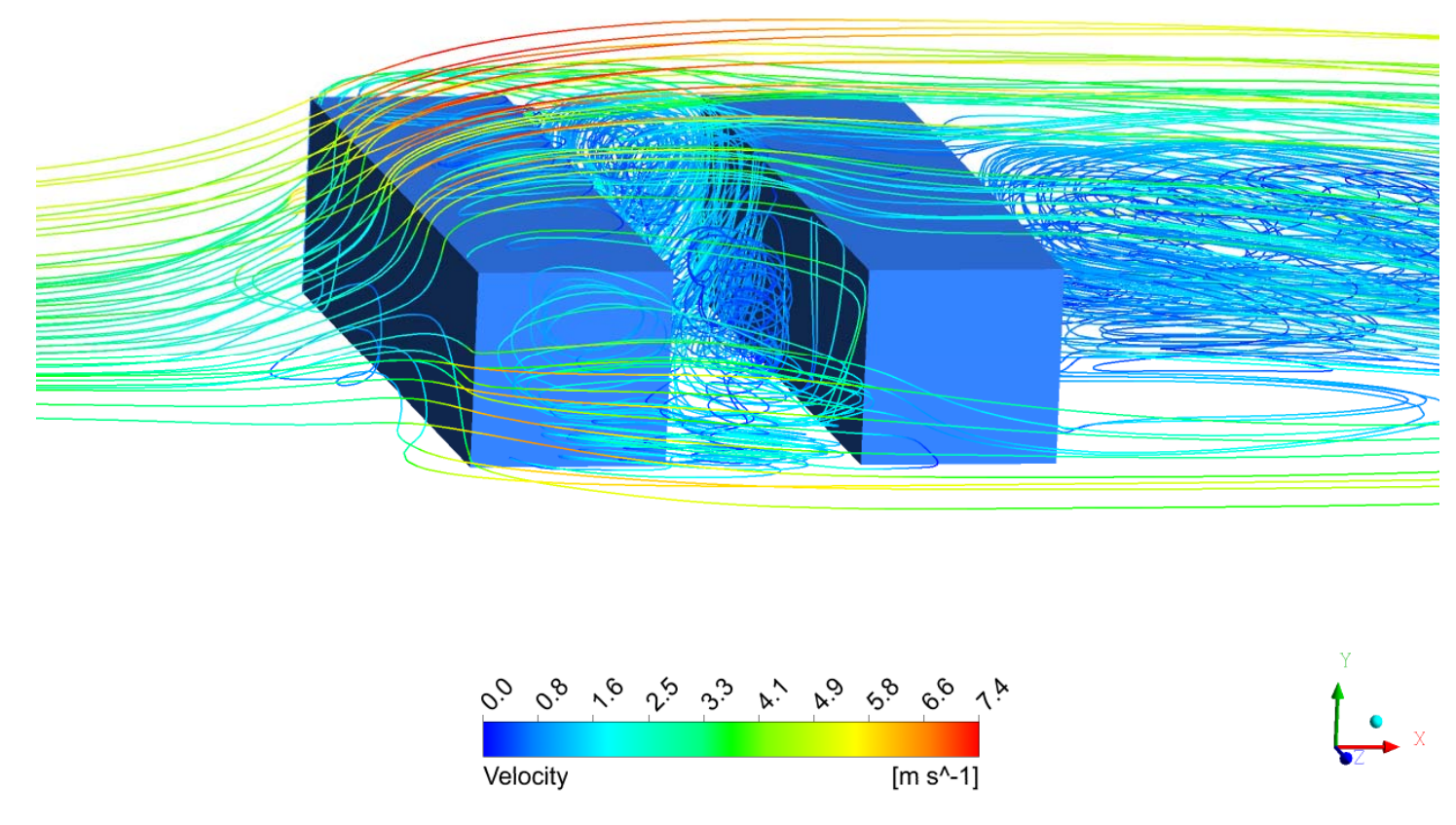}
	\vspace*{-3mm}
	\captionsetup[figure]{font=small,skip=0pt}
	\caption{Isometric view of streamlines showing coherent vortex structure around and inside the uniform street canyon for $S/H = 1$ and $W/H = 4$.}
	\label{Fig:4.9}
\end{figure}

\subsubsection{Boundary conditions}

We impose symmetry conditions at the top and lateral sides of the computational domain, that is we set the normal derivative of the velocity field equal to zero there. The outlet boundary condition was specified to be Fluent's ``outflow'' condition, in which the unknown properties at the outlet are extrapolated from the interior conditions (under certain smoothness and regularity assumptions satisfied here).  No-slip boundary conditions were set at the ground and at the building faces. In addition we introduce inflation layers in order to resolve any boundary layer flow;  $40$ grid layers was enough to accurately predict separation and reattachment points. The height of the first cell of the boundary layer was chosen to be $2.7\times10^{-5}$m to ensure the wall unit $y^{+}\approx1$ in order to resolve the viscous sub-layer of the boundary layer, a requirement of the turbulent model we use.  In this study the low-$Re$ (i.e. near-wall) modelling approach has been used to model the wall roughness effect. The boundary layer thickness based on the height of the building is $\delta/H\approx3.5$, which is the expected value for roughness length $y_{0}\approx 0.2$ m and corresponding power law exponent $\alpha=0.28$ for the real atmospheric boundary layer in which a power-law velocity profile is assumed.

\begin{figure}
	\vspace*{-2mm}
	\captionsetup[subfigure]{aboveskip=-2pt,belowskip=-2pt}
	\centering
	\begin{subfigure}[c]{.48\textwidth}
		\centering
		\includegraphics[width=1\linewidth]{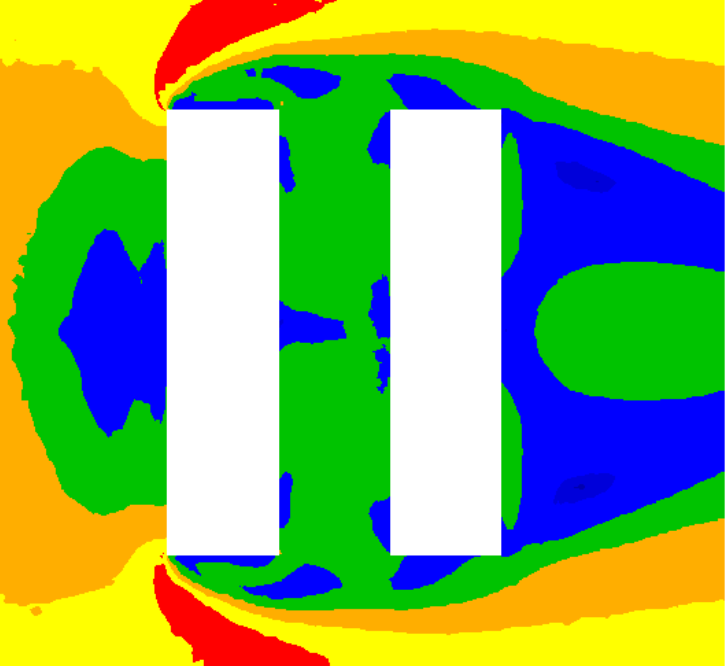}
		\caption{$S/H = 1$}
		\label{fig:sub4.10.4}
	\end{subfigure}\hfill
	\begin{subfigure}[c]{0.48\textwidth}
		\centering
		\includegraphics[width=1\linewidth]{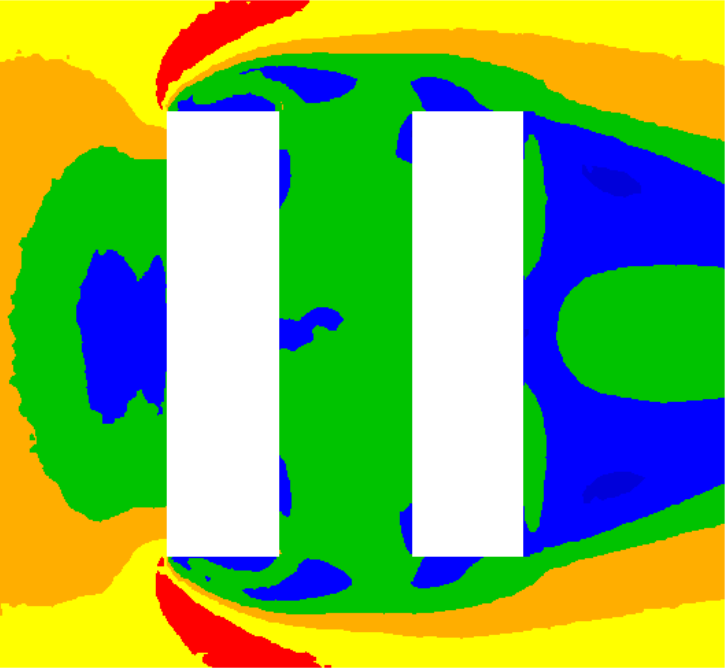}
		\caption{$S/H = 1.2$}
		\label{fig:sub4.10.5}
	\end{subfigure}\hfill
	\begin{subfigure}[c]{0.48\textwidth}
		\centering
		\includegraphics[width=1\linewidth]{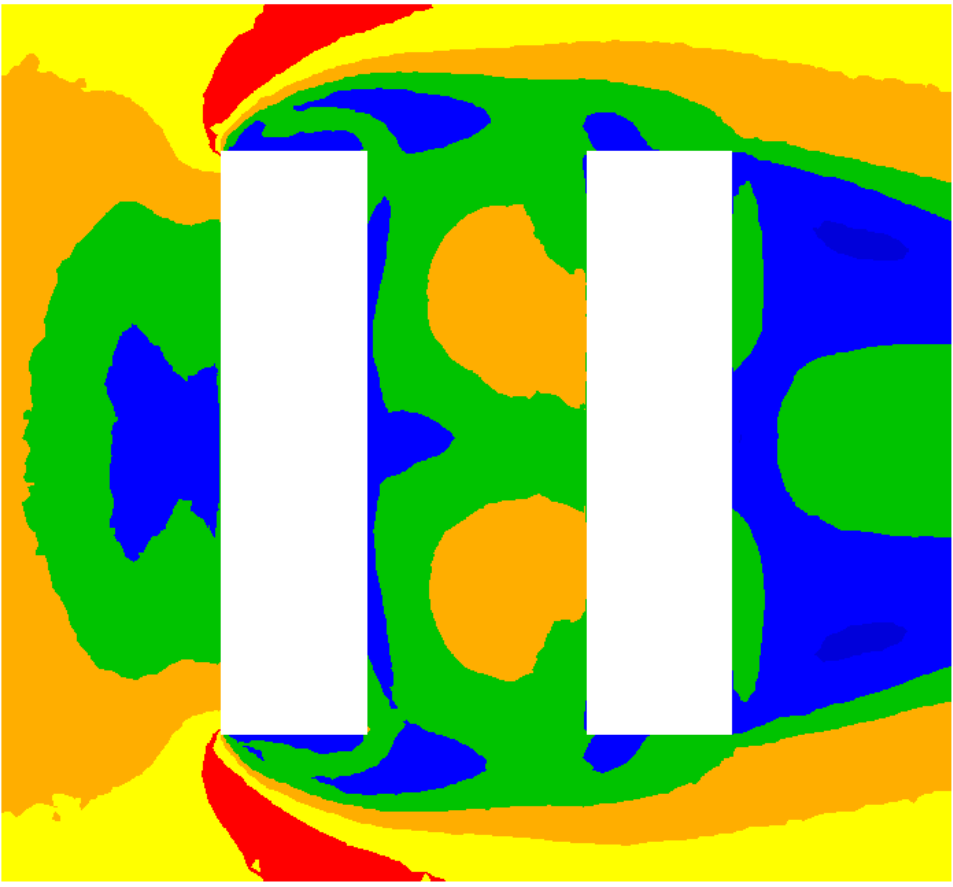}
		\caption{$S/H = 1.5$}
		\label{fig:sub4.10.6}
	\end{subfigure}\hfill
	\begin{subfigure}[c]{.48\textwidth}
		\centering
		\includegraphics[width=1\linewidth]{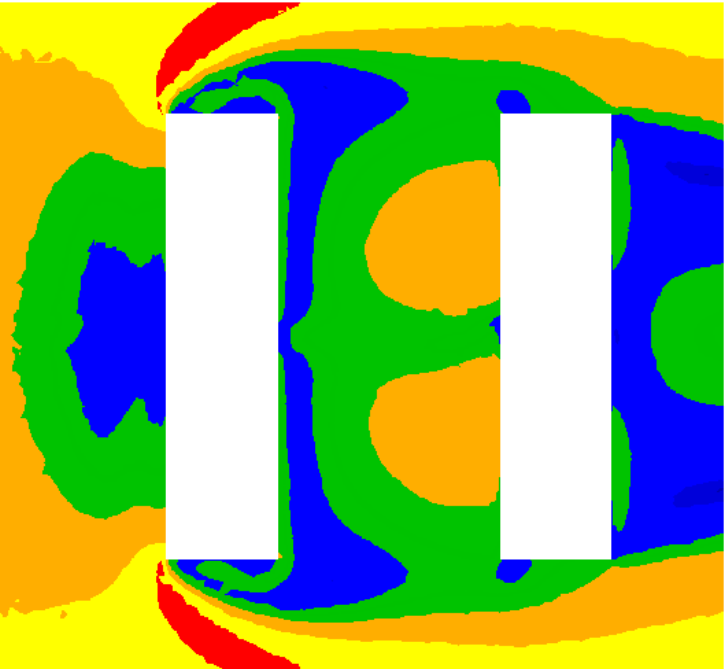}
		\caption{$S/H = 2$}
		\label{fig:sub4.10.7}
	\end{subfigure}\hfill
	\vspace*{-3mm}
	\captionsetup[figure]{font=small,skip=0pt}
	\caption{Wind comfort assessment at $Y = 1.5$m in the $XZ$-plane. As described in the text: blues and greens are comfortable, all other colours are uncomfortable.}
	\label{Fig:4.10}
\end{figure}

\subsection{Computational mesh}
  
We constructed a non-uniform mesh with tetrahedral and wedge-shaped elements. A mesh independence study in an empty channel (with roughness elements present but without buildings) was carried out to demonstrate the independence of the flow field with respect to mesh refinement. The coarse mesh had $3$ million cells of resolution $2.5$m throughout the computational domain. The medium mesh had $5.5$ million cells with resolution of $2$m throughout the computational domain and the fine mesh had $9.8$ million cells and a resolution of $1.6$m throughout the domain. The Reynolds number was set to be $8.1\times 10^{6}$ based on $y_{\text{ref}}=20$m and $U_{\text{ref}} =5.9$ m/s as mentioned above and as used in our canyon simulations reported in Sections \ref{sec:Results and discussion for uniform street canyons} and \ref{sec:Results and discussion for non-uniform street canyons}. The undisturbed vertical profiles of mean wind speed and turbulence measures were compared for all three meshes at the location where the building would be positioned. These profiles are called ``incident'' profiles and were in close agreement across all three meshes. In order to balance the time costs incurred by running multiple simulations across a variety of parameter values, and the desire for a reasonable resolution of the more complex flow patterns resulting from the presence of the street canyon, we used the medium mesh for all subsequent simulations presented herein. 

\begin{figure}
	\vspace*{-3mm}
	\captionsetup[subfigure]{aboveskip=-2pt,belowskip=-2pt}
	\centering
	\begin{subfigure}[c]{.48\textwidth}
		\centering
		\includegraphics[width=1\linewidth]{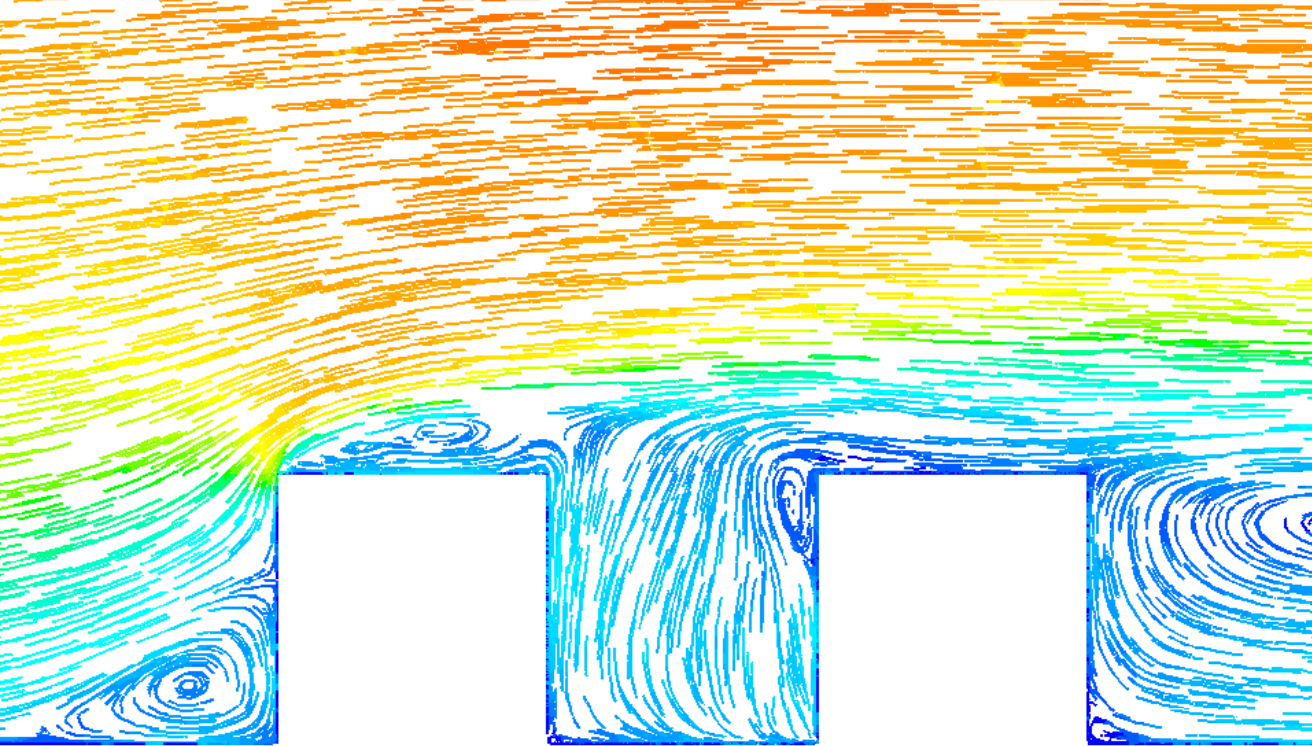}
		\caption{$W/H = 3$}
		\label{fig:sub4.11.1}
	\end{subfigure}
	\begin{subfigure}[c]{.48\textwidth}
		\centering
		\includegraphics[width=1\linewidth]{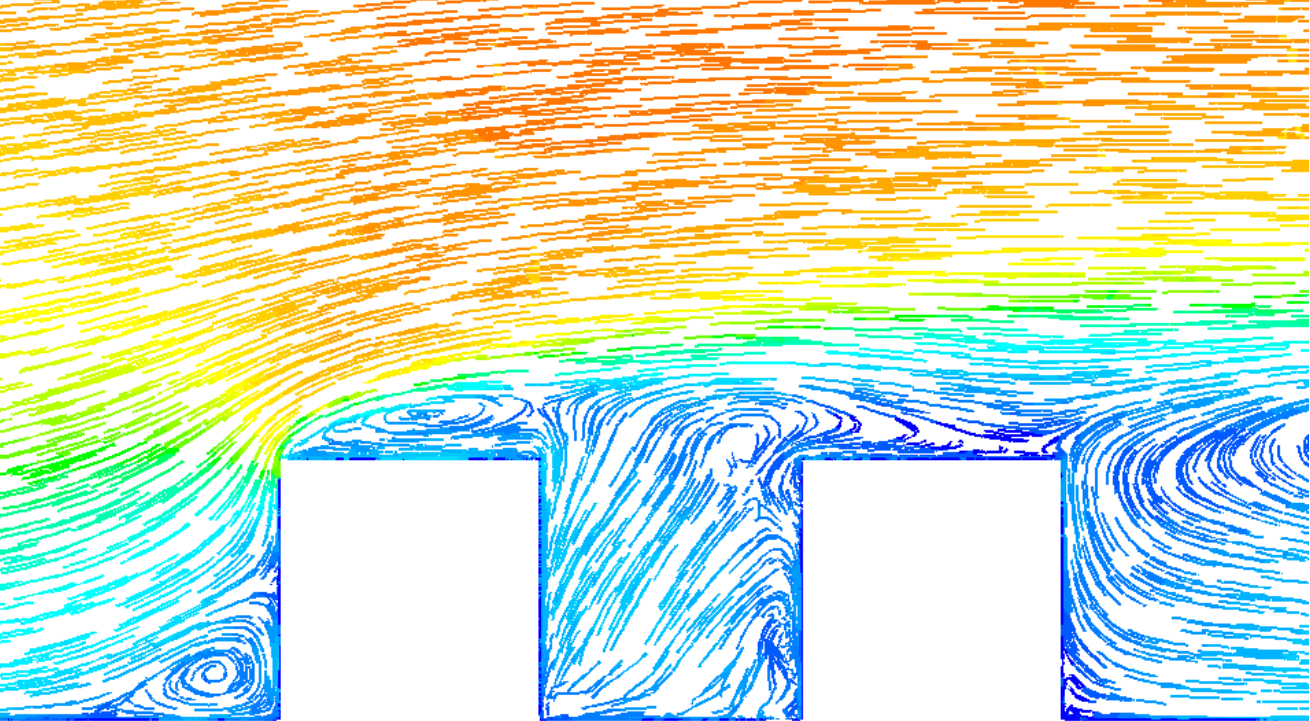}
		\caption{$W/H = 5$}
		\label{fig:sub4.11.2}
	\end{subfigure}
	\begin{subfigure}[c]{.48\textwidth}
		\centering
		\includegraphics[width=1\linewidth]{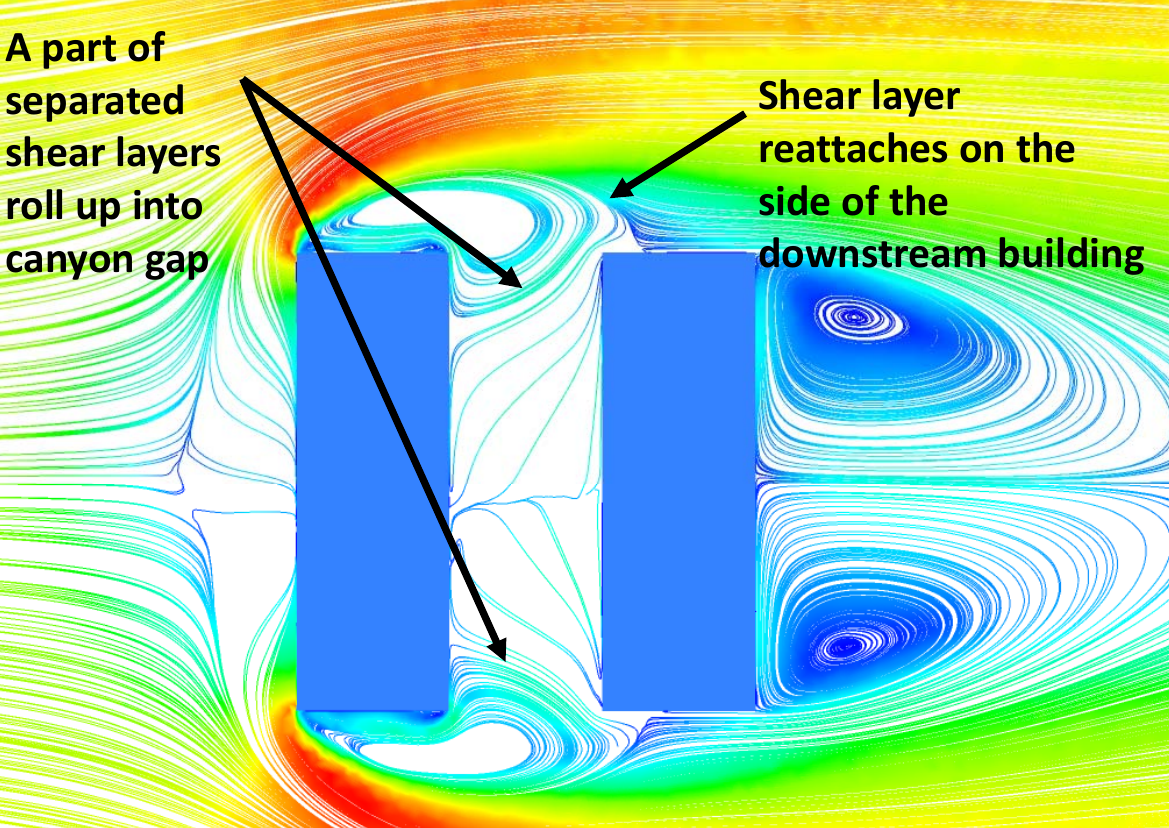}
		\caption{$W/H = 3$}
		\label{fig:sub4.11.3}
	\end{subfigure}
	\begin{subfigure}[c]{.48\textwidth}
		\centering
		\includegraphics[width=1\linewidth]{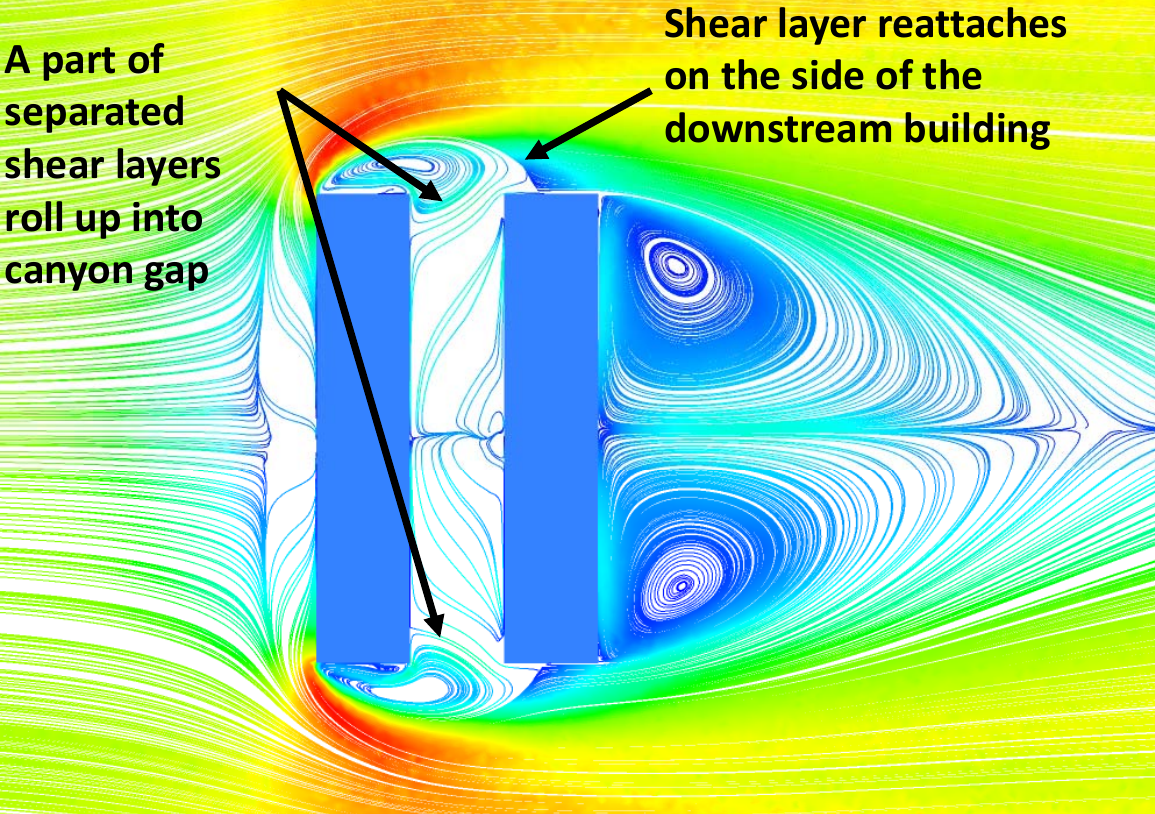}
		\caption{$W/H = 5$}
		\label{fig:sub4.11.4}
	\end{subfigure}
	\captionsetup[figure]{font=small,skip=0pt}
	\caption{Velocity magnitude pathlines in the $XY$-plane at the centre plane ($Z = 40$) for the uniform street canyon with different building widths and for the constant street aspect ratio $S/H = 1$.}
	\label{Fig:4.11}
\end{figure}

\subsection{Validation and turbulence model selection}

Turbulent flow around a surface-mounted three-dimensional obstacle is a standard benchmark problem, both experimentally and numerically. Much work has been done to understand the basic flow structure and dynamics of coherent vortex structures around buildings (e.g. \citet{key-10, key-26, key-20, key-35} and \citet{key-37}). Some of  the most detailed experimental measurements and flow visualizations were performed at Reynolds numbers of around 40--50,000 using our definition of $Re$, such as those by \citet{key-20}, \citet{key-25},and \citet{key-47}.
Another aim of this analysis is to select the best performing model. The performance of the turbulence models was also evaluated by comparison with a transient flow simulation using the Large Eddy Simulation (LES) approach.
%
%

In order to determine a suitable choice of turbulence model, we ran multiple simulations with different models and compared the results with experimental values from the literature. 
Figure~\ref{Fig:3.10} shows the comparison of the mean wind speed in the centre plane of the building. Overall all the steady-RANS turbulent models agree well with the experimental values. However, near the roof surface at $X/L=-0.25$, $U$ was negative in the experiment because of a reverse flow, which was only reproduced by the transition $k$-$kl$-$\omega$ model and by LES. We note that far downstream of the canyon there is a small discrepancy between simulated and measured values, but that this region is not our focus in the present study. Furthermore we note that while LES accurately reproduces the experimental results, it is considered to be too computationally expensive for the purposes of the present study, which requires many simulations to be run over a range of parameter values.

We also evaluated the performance of turbulence models in a horizontal plane near the ground. The transition $k$-$kl$-$\omega$ model predicted the experimental wind speed with an accuracy of around $85\%$ in the region with high wind speed where the wind speed ratio is $1$ or higher as shown in Figure~\ref{Fig:3.13}. Since this region is important in the evaluation of the pedestrian level wind environment, we use this turbulence model throughout.  

\begin{figure}
	\vspace*{-2mm}
	\captionsetup[subfigure]{aboveskip=-2pt,belowskip=-2pt}
	\centering
	\begin{subfigure}[c]{.48\textwidth}
		\centering
		\includegraphics[width=1\linewidth]{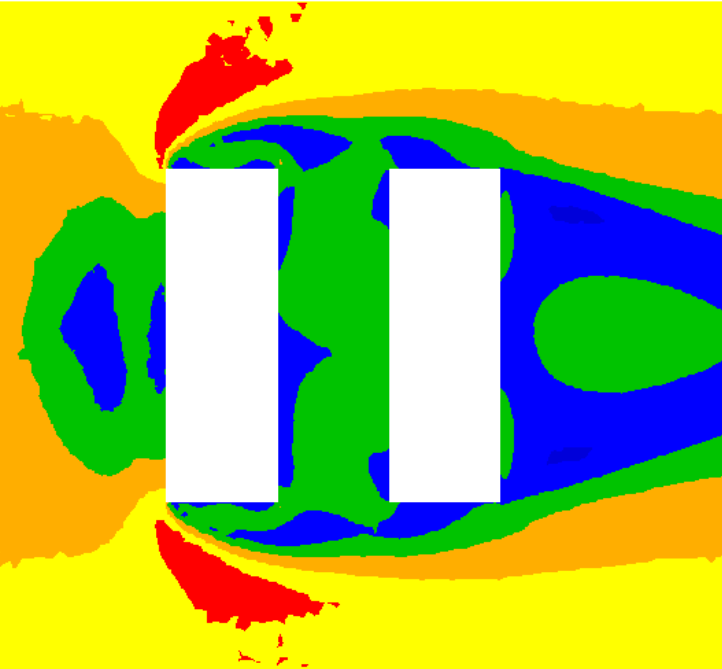}
		\caption{$W/H = 3$}
		\label{fig:sub4.12.1}
	\end{subfigure}
	\begin{subfigure}[c]{0.48\textwidth}
		\centering
		\includegraphics[width=1\linewidth]{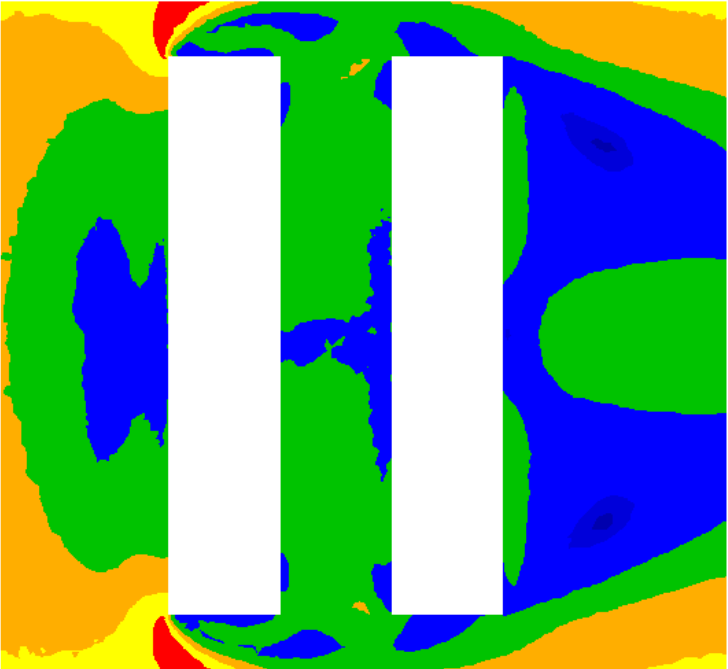}
		\caption{$W/H = 5$}
		\label{fig:sub4.12.2}
	\end{subfigure}	
	\vspace*{-3mm}
	\captionsetup[figure]{font=small,skip=0pt}
	\caption{Wind comfort assessment at $Y = 1.5$m in the $XZ$-plane for the uniform street canyon with different building widths and for the constant street aspect ratio $S/H = 1$. Colours as in Figure~\ref{Fig:4.10}.}
	\label{Fig:4.12}
\end{figure}

\subsection{Other parameters}

The Pressure-Implicit with Splitting of Operators (PISO) algorithm scheme with skewness correction was used for the pressure-velocity coupling; pressure interpolation was second-order. Second-order discretization schemes were used for both the convective terms and the viscous terms of the governing equations. The simulations were initialized with the values of the inlet boundary conditions. Surface monitor points inside the street canyon with $(X,Y,Z)$ coordinates $(22,8,40)$, $(26,12,30)$, $(30,7,20)$, $(34,5,70)$, and $(38,10,50)$ were used to measure convergence for the mean wind speed. (These are the points used for $S/H=1$ and $W/H=4$ and were changed accordingly with changes in $S/H$ and $W/H$.) The simulations were terminated when the residuals at all specified surface monitor points reached the convergence criterion of a difference in value between two iterations of $0.0005$ for $20$ consecutive iterations.

\begin{figure}
	\vspace*{-4mm}
	\captionsetup[subfigure]{aboveskip=-2pt,belowskip=-2pt}
	\centering
	\begin{subfigure}[b]{.43\textwidth}
		\centering
		\includegraphics[width=1\linewidth]{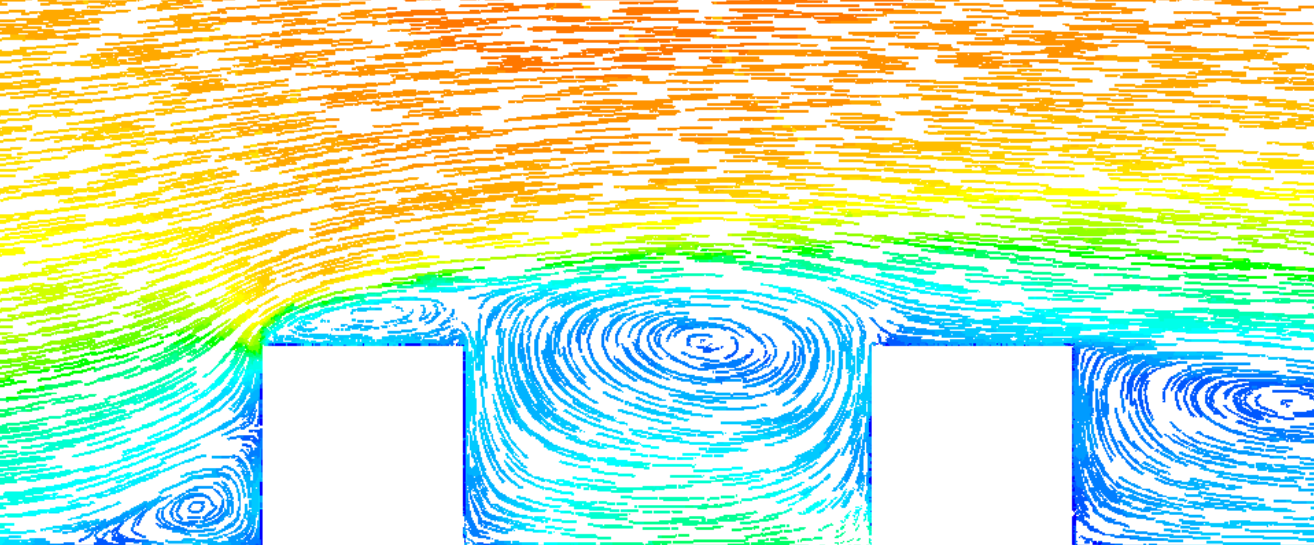}
		\caption{$W/H = 3$}
		\label{fig:sub4.13.1}
	\end{subfigure}
	\begin{subfigure}[b]{.4\textwidth}
		\centering
		\includegraphics[width=1\linewidth]{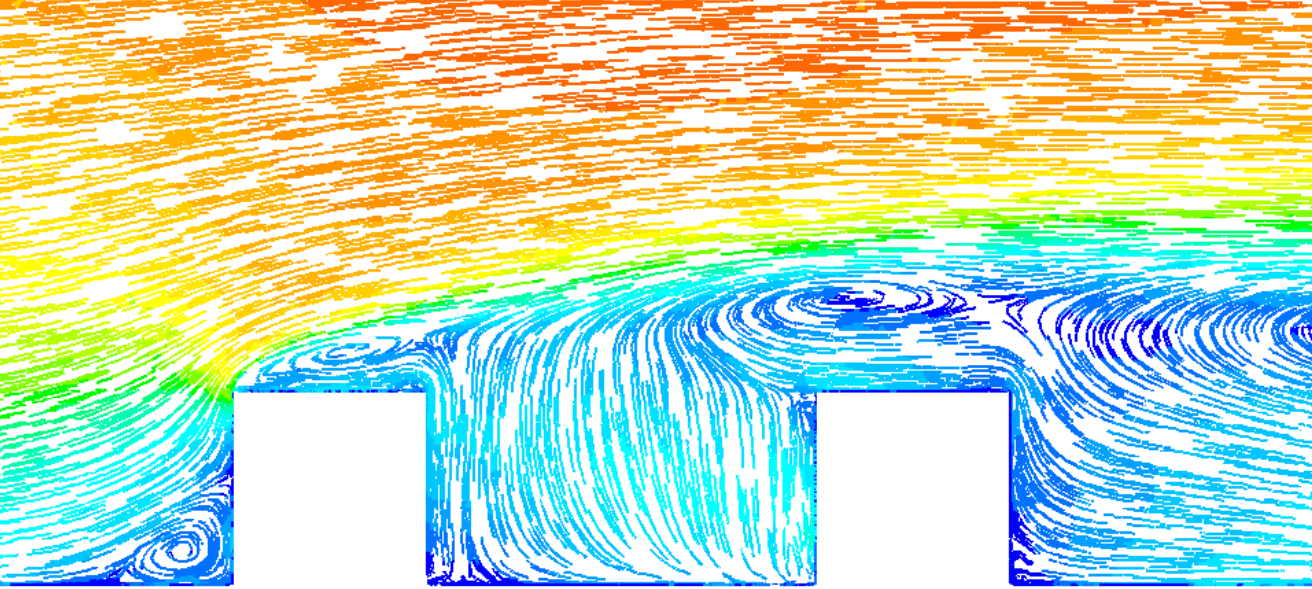}
		\caption{$W/H = 5$}
		\label{fig:sub4.13.2}
	\end{subfigure}
	\begin{subfigure}[b]{.42\textwidth}
		\centering
		\includegraphics[width=1\linewidth]{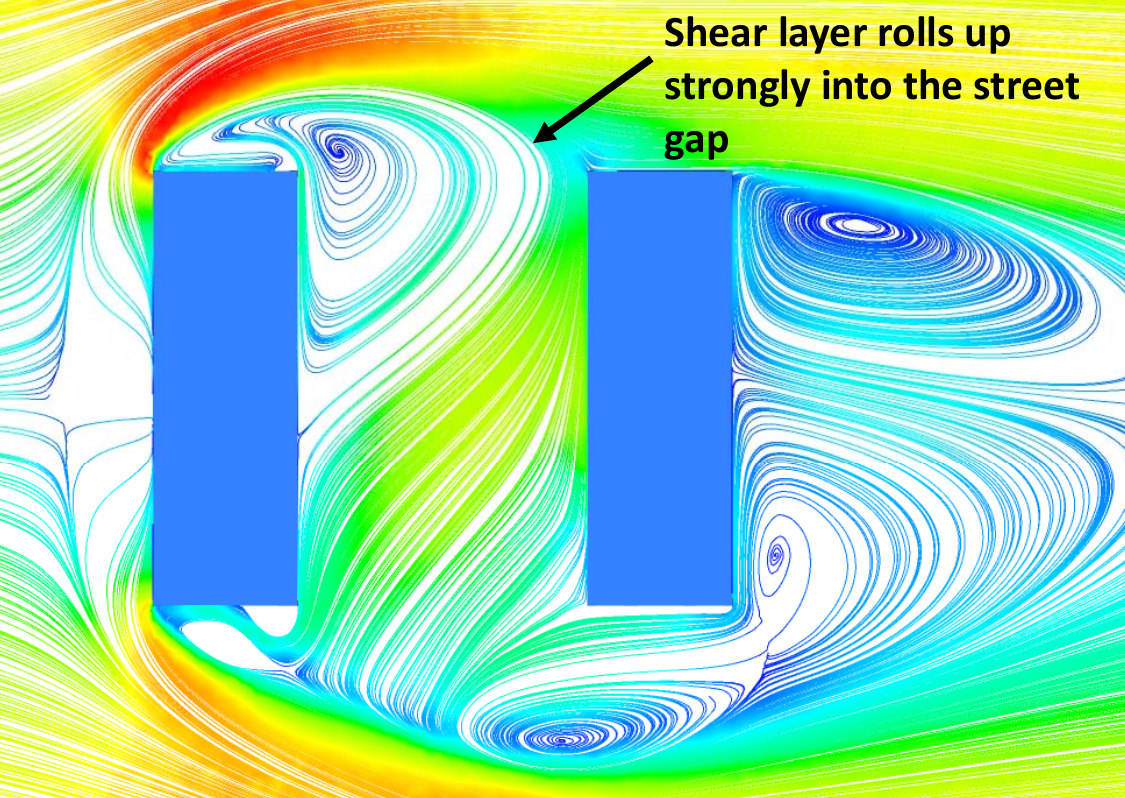}
		\caption{$W/H = 3$}
		\label{fig:sub4.13.3}
	\end{subfigure}
	\begin{subfigure}[b]{.42\textwidth}
		\centering
		\includegraphics[width=1\linewidth]{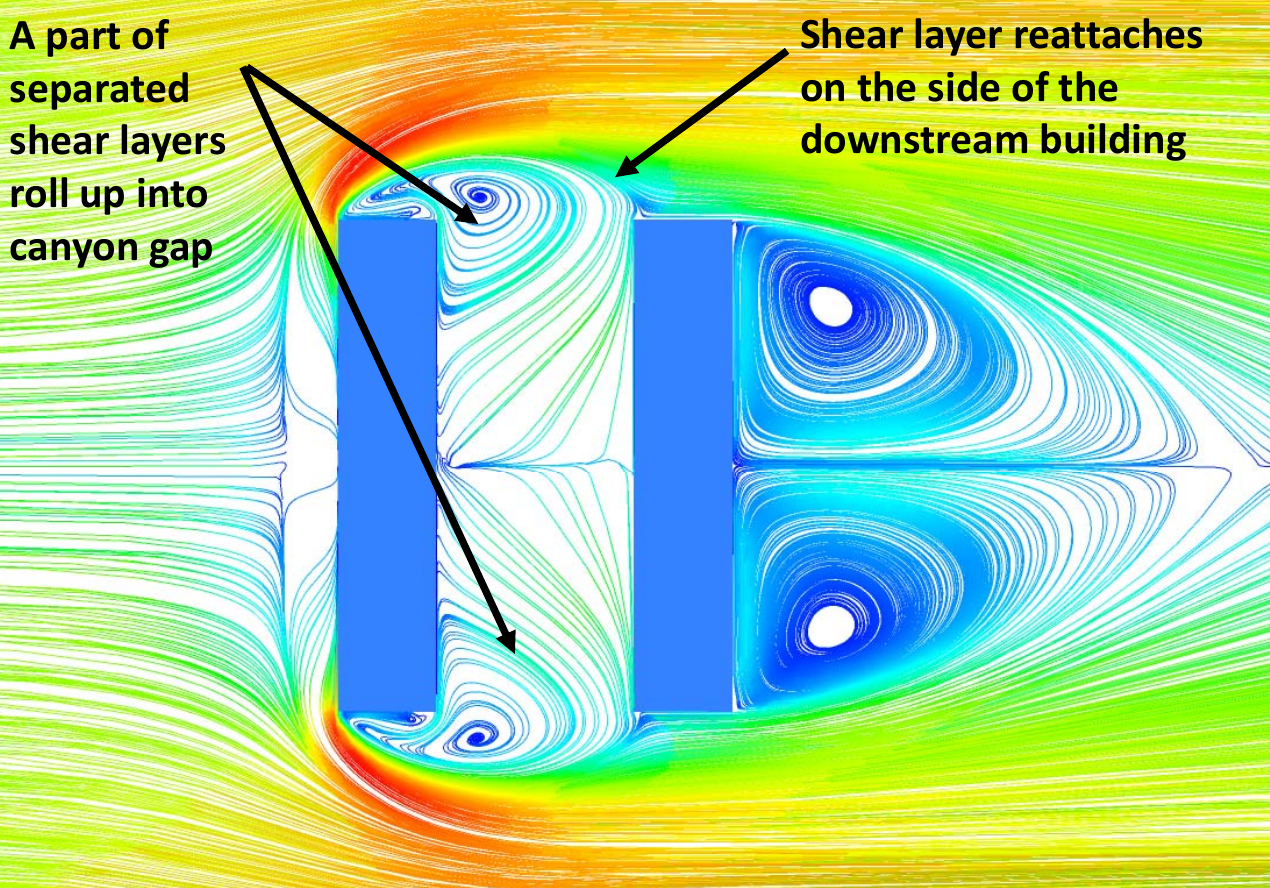}
		\caption{$W/H = 5$}
		\label{fig:sub4.13.4}
	\end{subfigure}
	\captionsetup[figure]{font=small,skip=0pt}
	\caption{Velocity magnitude pathlines in the $XY$-plane at the centre plane ($Z = 40$) for the uniform street canyon with different building widths and for the constant street aspect ratio $S/H = 2$.}
	\label{Fig:4.13}
\end{figure}

%
%

\section{Results: uniform canyons}
\label{sec:Results and discussion for uniform street canyons}

\subsection{Avenue and Regular Canyons of short size}
\label{subsec:Street width as an influencing parameter}

In this section we consider $S/H = 1$, $1.2$, $1.5$ and $2$ and $S/W=1/4$. For all our simulations the Reynolds number was $8.1 \times 10^{6}$ based on the reference height $y_{\text{ref}}=20$m and corresponding wind speed $U_{\text{ref}}=5.9$ m/s. This choice of reference scales gives us a wind speed corresponding to an ELBS number of $3$ at pedestrian height of approximately $1.75$m. Finally, we fix $W/H=4$, so we are in the short canyon range, and consider the effects of varying this parameter in the subsequent subsection.

As we can see in Figure \ref{Fig:4.6} some of the flow characteristics at the upstream building are quite similar to the single obstacle problem for all the different canyons. The flow separates upstream of the windward face of the obstacle with the formation of a stagnation point, which for all cases occurs at about $0.6 H$ from the ground. At this point the flow deviates into three main streams that can be classified as follows: (i) a recirculation region below the stagnation point, see Figure \ref{fig:sub4.7.2}; (ii) flow separation due to the sharp windward edges of the building which reattaches on the roof and then bifurcates, see Figure \ref{Fig:4.6}; and (iii) on the side walls of the building a horseshoe vortex forms and carries the surrounding fluid into the centre of the canyon, see Figure \ref{Fig:4.8}. All of these features can also be seen in the isometric view in Figure \ref{Fig:4.9}.

The reattachment of the shear-layer on the upwind building roof is validated by the results of \citet{key-10}, which showed that if $\delta/H > 0.7$ the separated shear layer permanently reattaches on the top of the building; for all our cases we have $\delta/H \approx 3.5$. The separated flow on the roof gives rise to a large recirculation region and backflow from the canyon into the roof is present --- see Figure \ref{fig:sub4.7.2}. 

On the other hand, the flow dynamics on the roof of the downstream building are dependent on the ratio $S/H$. The flow from inside the canyon streams upwards and towards the windward face of the second building and a strong vortex develops on its roof. The size of the vortex increases as $S/H$ increases. We note that our results are qualitatively in accordance with \citet{key-37}, in which if $S/H <2$ the separated shear layers from both sides of the upstream building attached to the side surfaces of the downstream building and no periodicity was seen in the wake. The authors called this the \emph{stable reattachment regime}. This regime is clearly seen in Figure~\ref{Fig:4.8}(A), Figure~\ref{Fig:4.8}(B) and Figure~\ref{Fig:4.8}(C). For the case $S/H = 2$ shown in Figure~\ref{Fig:4.8}(D) (and, indeed, for $S/H \geqslant 2$), a part of the separated shear layers from the upstream building rolls up intermittently into the canyon and periodicity is observed in the wake. \citet{key-37} referred to this as the \emph{unstable reattachment regime} or \emph{bistable flow}.

In Figure~\ref{Fig:4.10} we present a wind comfort chart based on the ELBS number. The contours are divided into six colours. Those denoting comfortable regions are dark blue, blue, and green, defined as: dark blue (wind speeds of 0--0.1m/s, ELBS number 0); blue (0.2--1.0m/s, ELBS 1); and green (1.1--2.3m/s, ELBS 2). The uncomfortable regions are coloured orange, yellow, and red: orange (2.4--3.8m/s, ELBS 3); yellow (3.9--5.5m/s, ELBS 4); and red (5.6--7.5m/s, ELBS 5). Clearly, regular canyons are well within the comfortable zone. They are perfect for pedestrian boulevards and outside sitting; see Figures \ref{Fig:4.10} A and B. On the other hand, avenue canyons are prone to uncomfortable regions, specifically on the upwind side of the downwind building; see Figures \ref{Fig:4.10} C and D. This was to be expected because as $S/H$ increases the separated shear layers from the sides of the upstream building roll up strongly into the gap between the buildings, as a consequence more fluid enters the canyon and we have an increase in the wind speed.


\begin{figure}
	\vspace*{-3mm}
	\captionsetup[subfigure]{aboveskip=-2pt,belowskip=-2pt}
	\centering
	\begin{subfigure}[b]{.4\textwidth}
		\centering
		\includegraphics[width=1\linewidth]{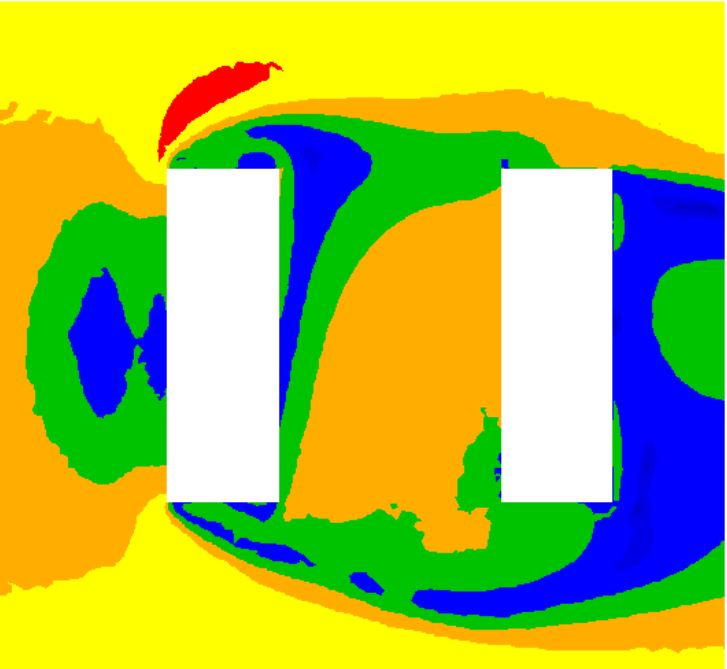}
		\caption{$W/H = 3$}
		\label{fig:sub4.14.1}
	\end{subfigure}
	\begin{subfigure}[b]{0.4\textwidth}
		\centering
		\includegraphics[width=1\linewidth]{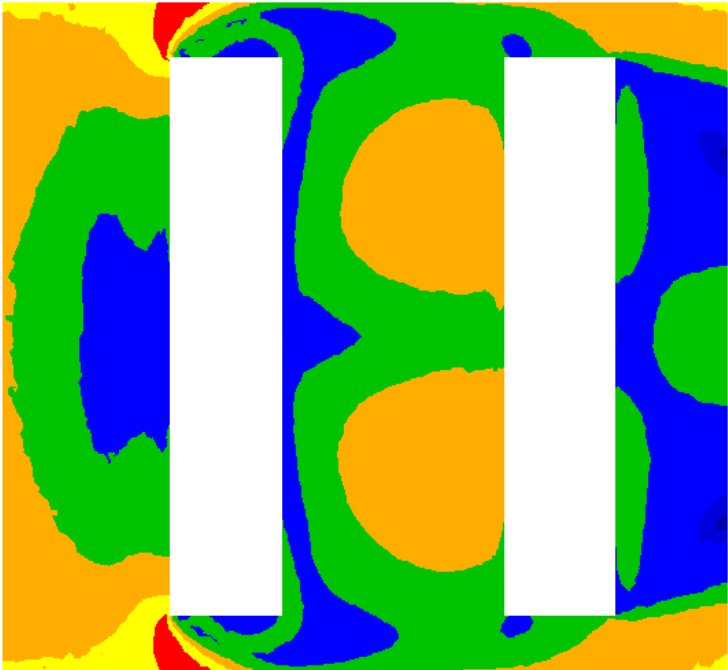}
		\caption{$W/H = 5$}
		\label{fig:sub4.14.2}
	\end{subfigure}
	\vspace*{-3mm}
	\captionsetup[figure]{font=small,skip=0pt}
	\caption{Wind comfort assessment at $Y = 1.5$m in the $XZ$-plane for the uniform street canyon with different building width and for the constant street aspect ratio $S/H = 2$.  Colours as in Figure~\ref{Fig:4.10}.}
	\label{Fig:4.14}
\end{figure}

%
%

\subsection{Effects of increasing or decreasing building width on pedestrian comfort}
\label{subsec:Building width as an influencing parameter-partone}

In this section we explore the effect that varying the width of the buildings has on the flow characteristics by means of increasing or decreasing the ratio $W/H$. We choose to investigate only the cases $W/H=3$, which gives a standard small canyon and $W/H=5$ which represents a medium canyon. We present results for the cases $S/H=1$ (shown to be predominantly comfortable for pedestrians) and $S/H=2$ (which showed a large uncomfortable zone).

In Figure \ref{Fig:4.11} we present the velocity field at the centre of the canyon and the flow streamlines at height 1.5 m. Clearly, the ratio $W/H$ has a negligible effect on the flow and our results are qualitatively the same as the ones presented in section \ref{subsec:Street width as an influencing parameter}. This is further reinforced by the ELBS charts in Figure~\ref{Fig:4.12}. As before, the maximum Beaufort number inside the canyon is 2.  

In Figure \ref{Fig:4.13} we can see that even though $S/H=2$, for the case the case $W/H=3$, the flow is no longer bistable (as it was in section \ref{subsec:Street width as an influencing parameter}). It has undergone a bifurcation and is now in what \citet{key-27} call a \emph{lock in} regime, in which the upstream shear layer impinges on the leading face (near the leading edge) of the downstream building and a strong vortex rolls in the canyon. This is clearly seen in Figures \ref{Fig:4.13} (A) and (C). As $W/H$ increases we move back to the bistable regime and stay there for the case $W/H=5$ --- see Figures \ref{Fig:4.13} (B) and (D).

Figure~\ref{Fig:4.14} shows the pedestrian wind comfort assessment plot for varying building width for $S/H = 2$ at the pedestrian height of $1.5$ m. As expected, for the case of a small avenue canyon we have an ELBS number 3 region (in orange) occupying a significant region of the canyon in both cases. Here, an increase of the building width does not seem to strongly affect the wind speed inside the canyon.

\begin{figure}[t]
	\vspace*{-5mm}
	\captionsetup[subfigure]{aboveskip=-2pt,belowskip=-2pt}
	\centering
	\begin{subfigure}[c]{.48\textwidth}
		\centering
		\includegraphics[width=1\linewidth]{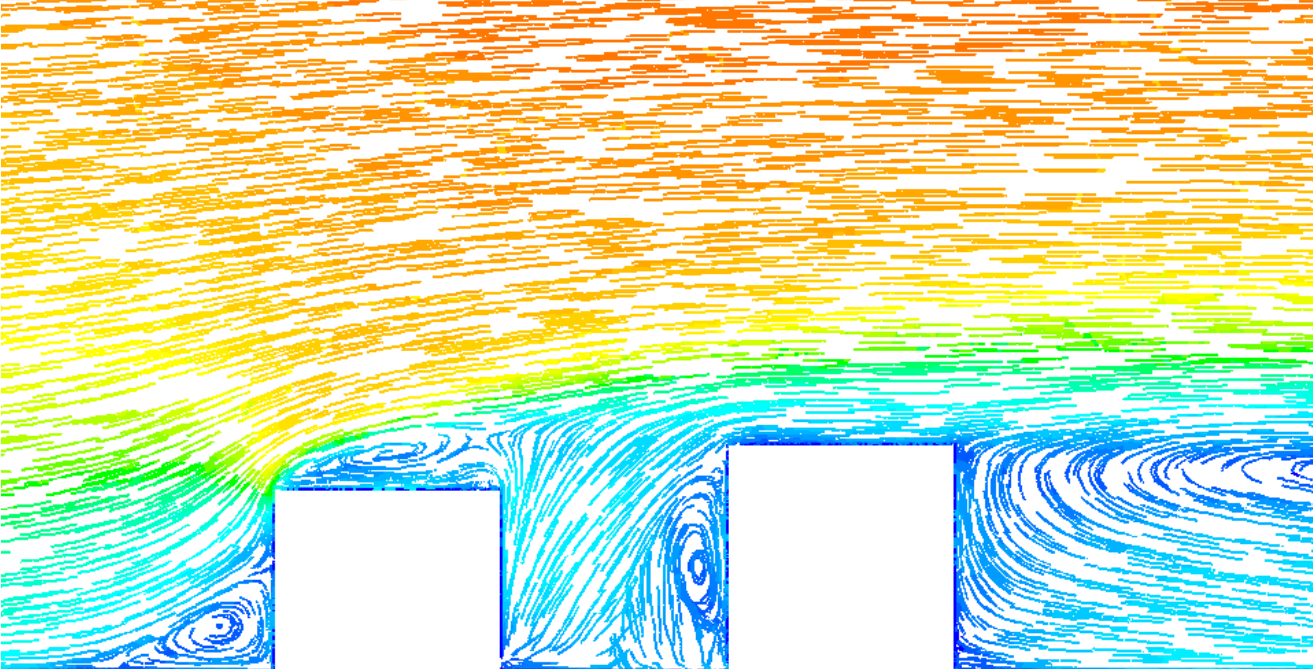}
		\caption{Step-up SC $S/H = 1$}
		\label{fig:sub4.17.2}
	\end{subfigure}\hfill
	\begin{subfigure}[c]{.48\textwidth}
		\centering
		\includegraphics[width=1\linewidth]{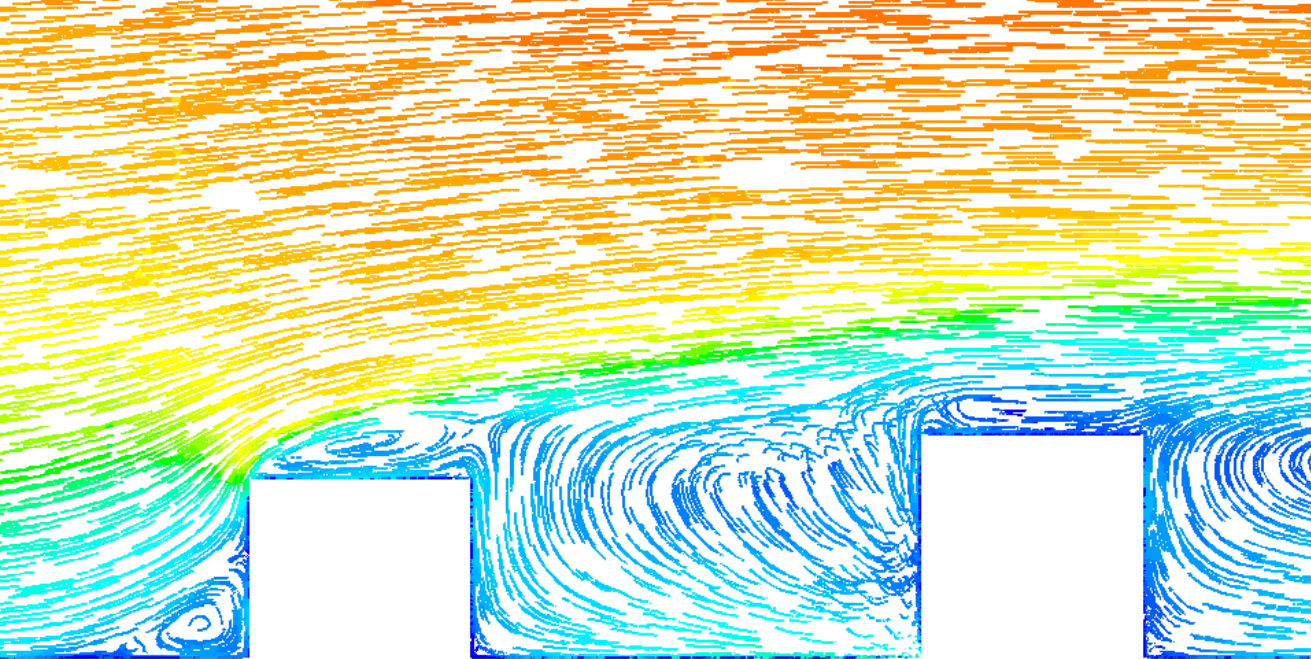}
		\caption{Step-up SC $S/H = 2$}
		\label{fig:sub4.17.3}
	\end{subfigure}\hfill
	\begin{subfigure}[c]{.48\textwidth}
		\centering
		\includegraphics[width=1\linewidth]{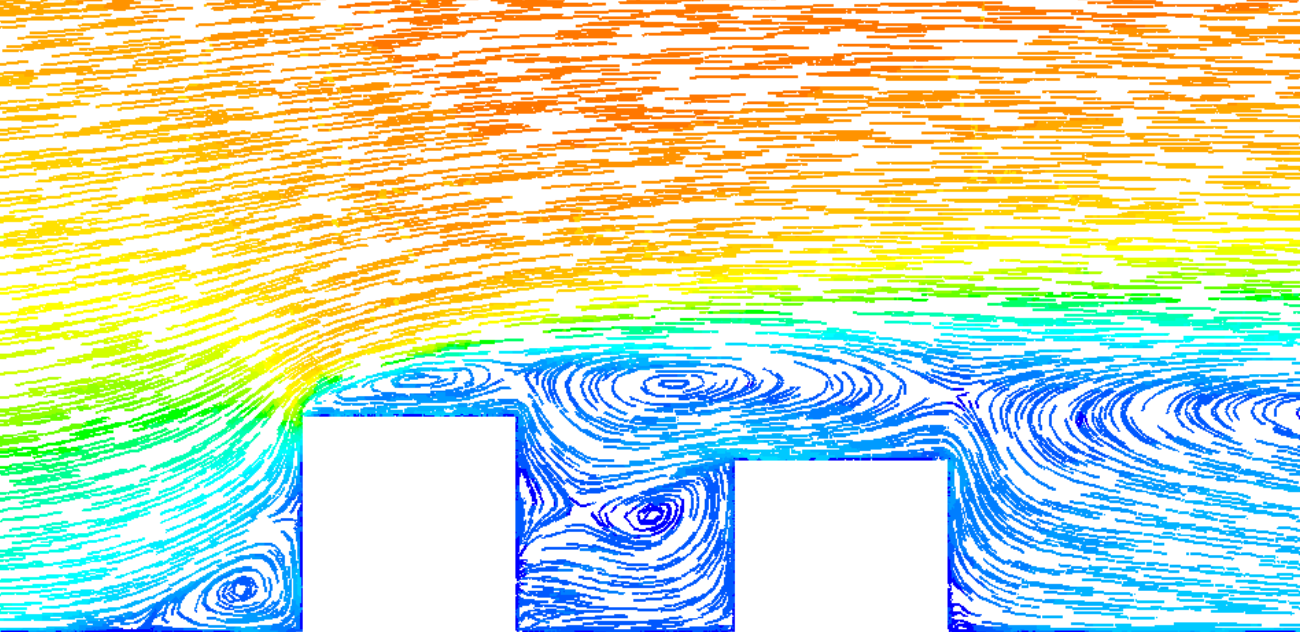}
		\caption{Step-down SC $S/H = 1$}
		\label{fig:sub4.17.4}
	\end{subfigure}\hfill
	\begin{subfigure}[c]{0.48\textwidth}
		\centering
		\includegraphics[width=1\linewidth]{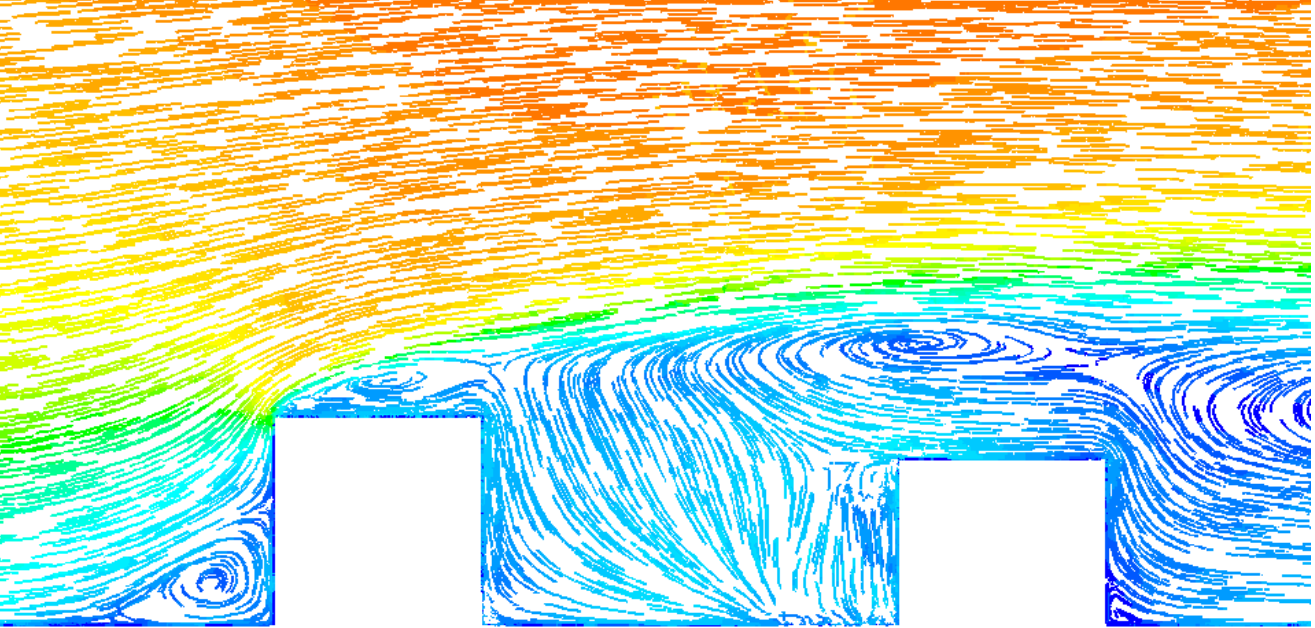}
		\caption{Step-down SC $S/H = 2$}
		\label{fig:sub4.17.5}
	\end{subfigure}
	\vspace*{-3mm}
	\captionsetup[figure]{font=small,skip=0pt}
	\caption{Velocity magnitude pathlines in the $XY$-plane at the centre plane ($Z = 40$) for the non-uniform street canyon cases.}
	\label{Fig:4.17}
\end{figure}

\begin{figure}[t]
	\vspace*{-3mm}
	\captionsetup[subfigure]{aboveskip=-2pt,belowskip=-2pt}
	\centering
	\begin{subfigure}[c]{.48\textwidth}
		\centering
		\includegraphics[width=1\linewidth]{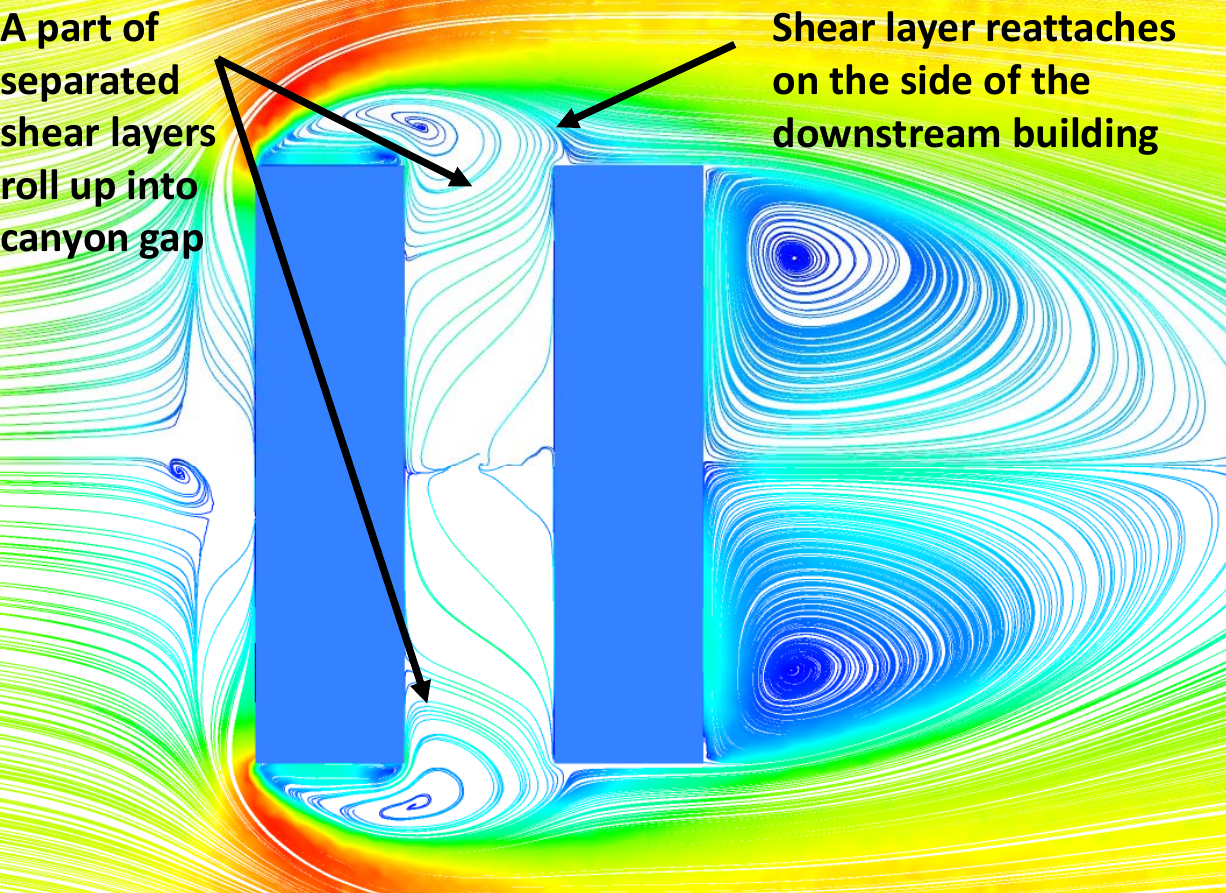}
		\caption{step-up SC $S/H = 1$}
		\label{fig:sub4.18.2}
	\end{subfigure}
	\begin{subfigure}[c]{.48\textwidth}
		\centering
		\includegraphics[width=1\linewidth]{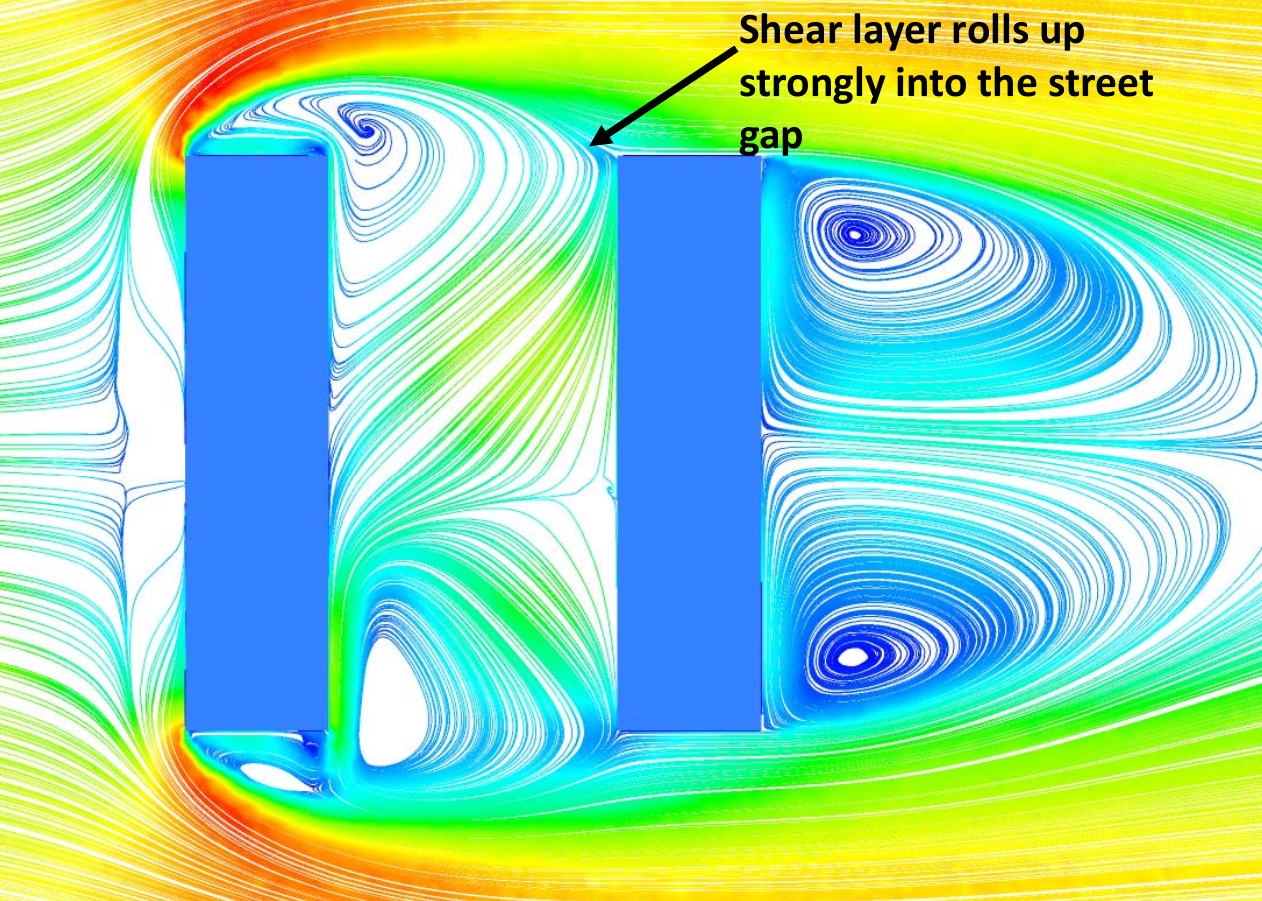}
		\caption{step-up SC $S/H = 2$}
		\label{fig:sub4.18.3}
	\end{subfigure}
	\begin{subfigure}[c]{.48\textwidth}
		\centering
		\includegraphics[width=1\linewidth]{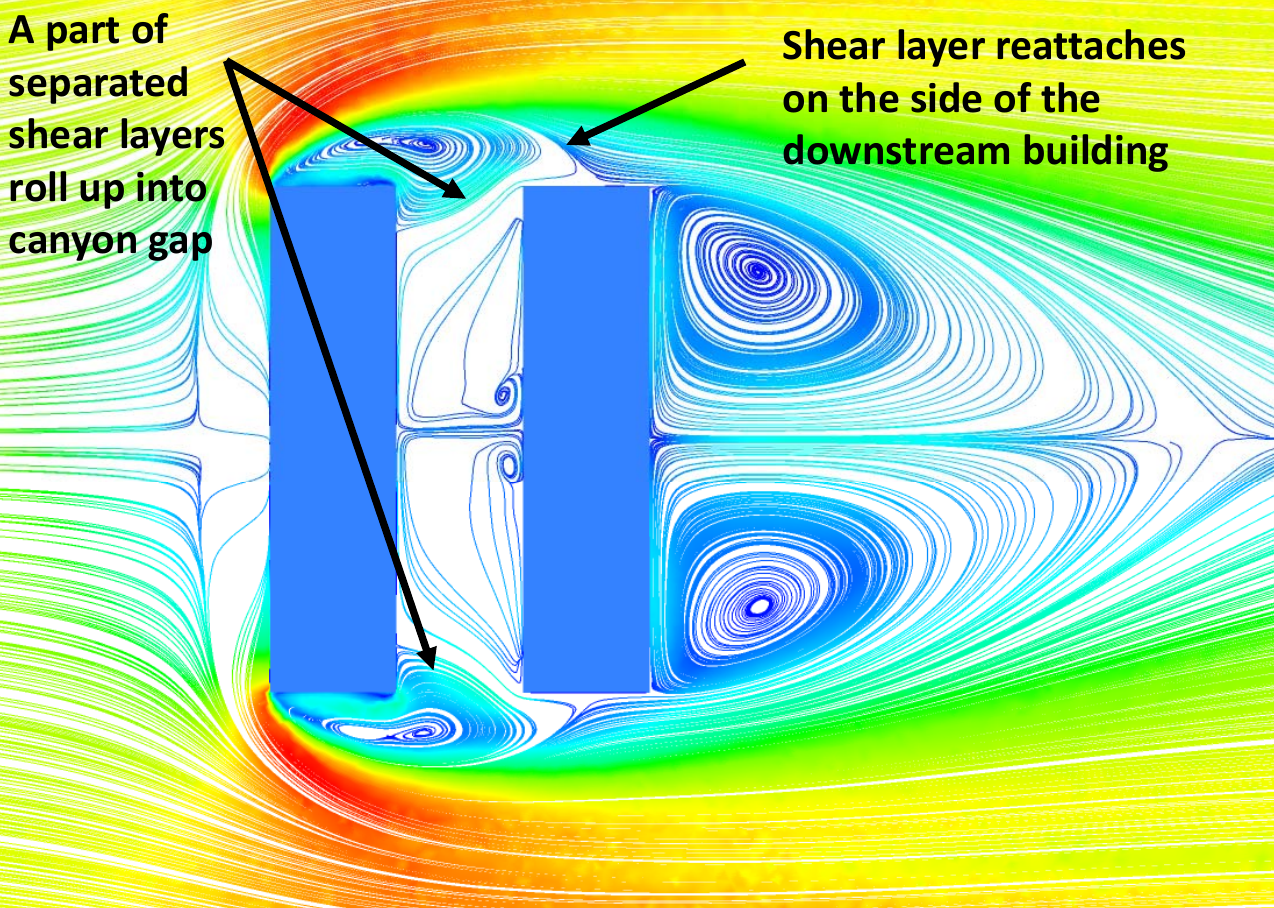}
		\caption{step-down SC $S/H = 1$}
		\label{fig:sub4.18.4}
	\end{subfigure}
	\begin{subfigure}[c]{0.48\textwidth}
		\centering
		\includegraphics[width=1\linewidth]{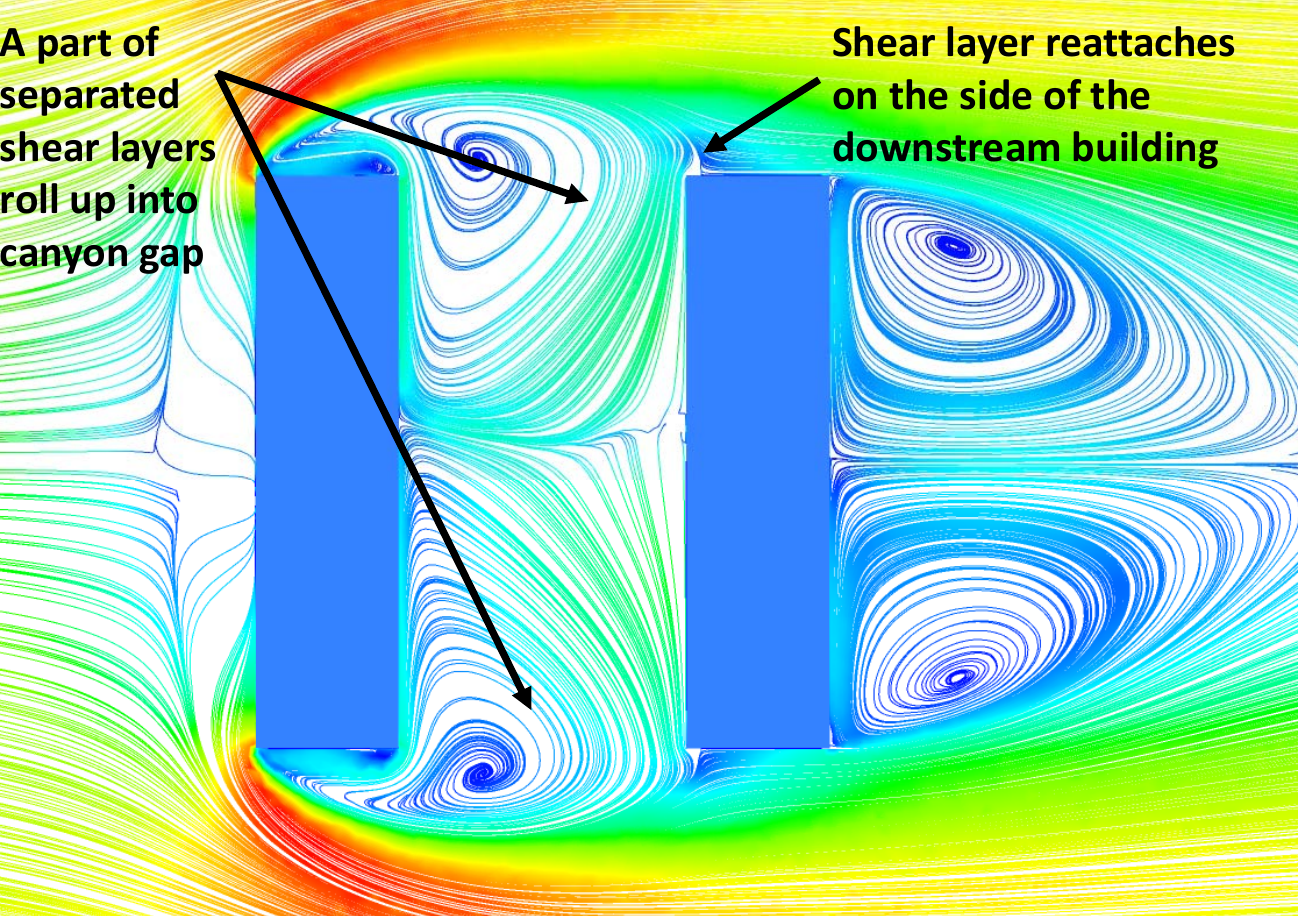}
		\caption{step-down SC $S/H = 2$}
		\label{fig:sub4.18.5}
	\end{subfigure}
	\vspace*{-3mm}
	\captionsetup[figure]{font=small,skip=0pt}
	\caption{Velocity magnitude streamlines in the $XZ$-plane at $Y = 1.5$m for the non-uniform street canyon cases.}
	\label{Fig:4.18}
\end{figure}

%
%

\section{Results:  non-uniform street canyons}
\label{sec:Results and discussion for non-uniform street canyons}

In this section we present our results for the case of non-uniform canyons. As shown in Section \ref{subsec:Building width as an influencing parameter-partone}, the effect of building width on the ELBS number inside the canyon is negligible. Hence we will fix $W/H=4$ and consider the cases $S/H = 1$ (seen to be favourable for pedestrian comfort in Section~\ref{sec:Results and discussion for uniform street canyons}) and $S/H = 2$ (generally unfavourable) for this analysis. Also, note that $H=H_L$ where $H_L$ is the height of the largest building (20 m). We set the height of the shortest building as $H_S=0.8H_L$. Therefore, for a step-up canyon we have $H_1=H_S$ and $H_2=H_L$ and for a step-down canyon the values are reversed.

Most of the flow characteristics are similar to those discussed in subsection~\ref{subsec:Street width as an influencing parameter} for the uniform street canyon. 
However, there are some key differences in the flow structure which have measurable effects on pedestrian comfort.     

Figure~\ref{Fig:4.17} shows results for the flow structure in the centre plane for four cases. For the step-up street canyon cases, the amount of fluid entering the canyon from the roof of the upstream building is larger than the corresponding uniform street canyon cases, as is clearly shown in Figures~\ref{fig:sub4.17.2} and ~\ref{fig:sub4.17.3}.

For the step-down street canyon cases, a large back-flow or a secondary re-circulation region can be observed on the roof of the downstream building in Figures~\ref{fig:sub4.17.4} and ~\ref{fig:sub4.17.5} respectively. This secondary recirculation was not seen in any other cases herein.

\begin{figure}[t]
	\captionsetup[subfigure]{aboveskip=-2pt,belowskip=-2pt}
	\centering
	\begin{subfigure}[c]{.48\textwidth}
		\centering
		\includegraphics[width=1\linewidth]{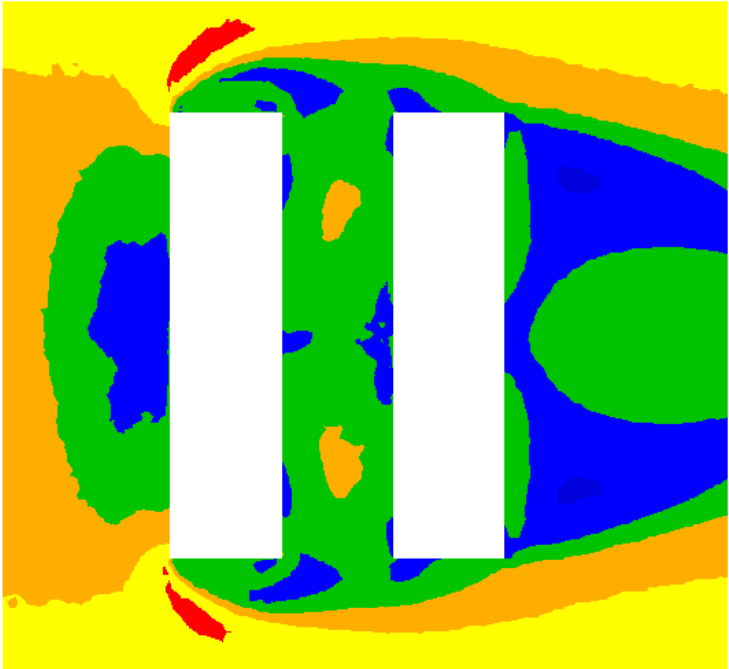}
		\caption{Step-up SC $S/H = 1$}
		\label{fig:sub4.19.2}
	\end{subfigure}\hfill
	\begin{subfigure}[c]{0.48\textwidth}
		\centering
		\includegraphics[width=1\linewidth]{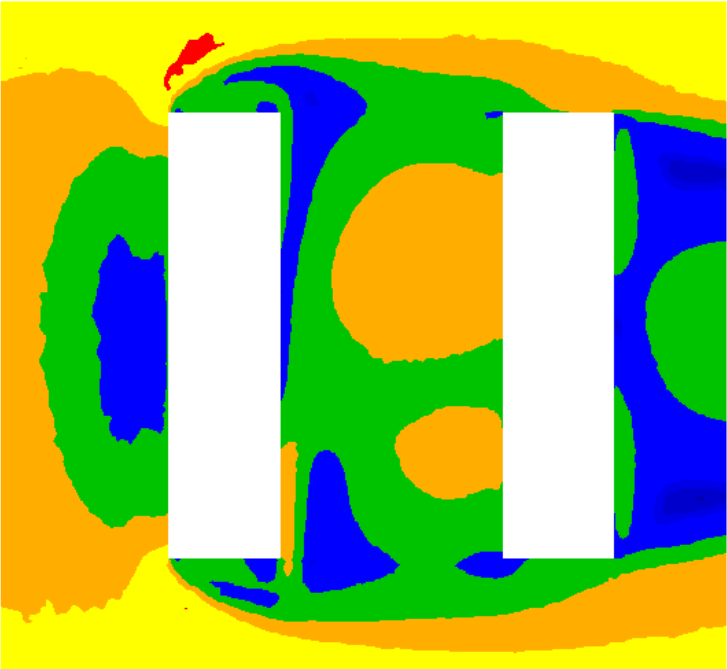}
		\caption{Step-up SC $S/H = 2$}
		\label{fig:sub4.19.3}
	\end{subfigure}\hfill
	\begin{subfigure}[c]{.48\textwidth}
		\centering
		\includegraphics[width=1\linewidth]{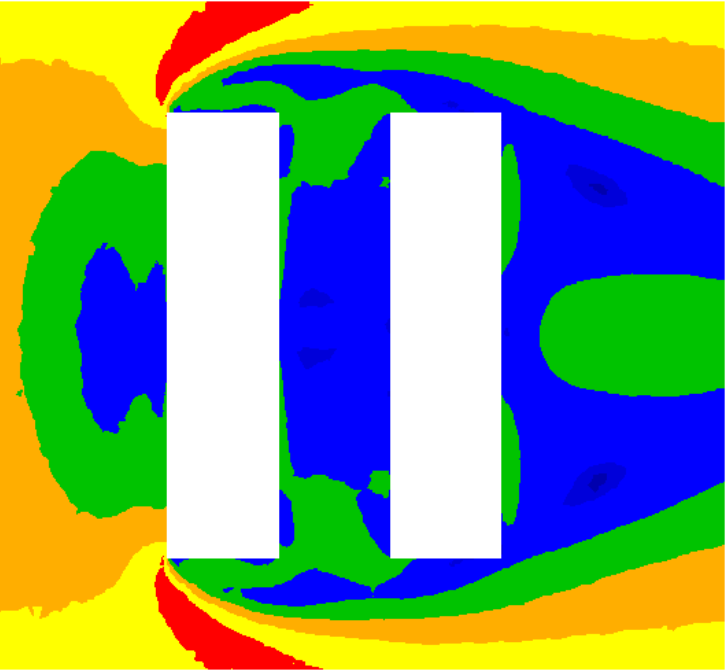}
		\caption{Step-down SC $S/H = 1$}
		\label{fig:sub4.19.4}
	\end{subfigure}\hfill
	\begin{subfigure}[c]{0.48\textwidth}
		\centering
		\includegraphics[width=1\linewidth]{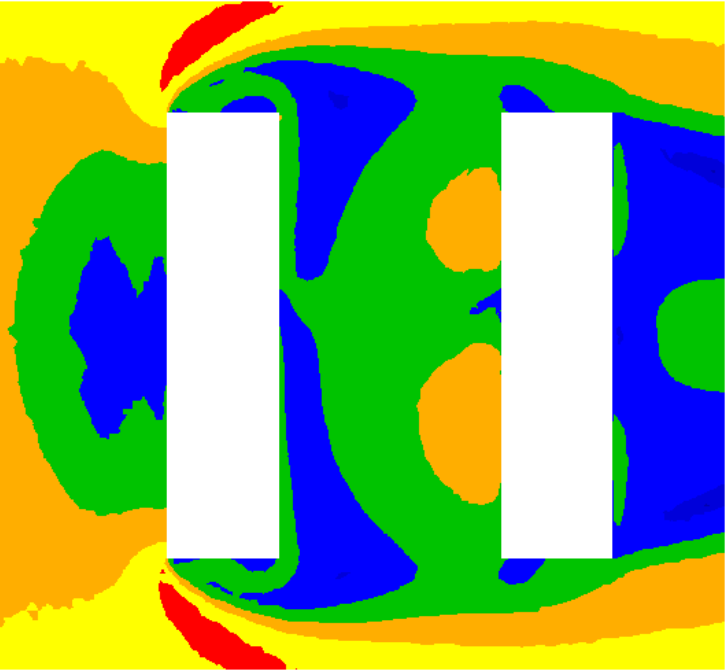}
		\caption{Step-down SC $S/H = 2$}
		\label{fig:sub4.19.5}
	\end{subfigure}\hfill
	\vspace*{-3mm}
	\captionsetup[figure]{font=small,skip=0pt}
	\caption{Wind comfort assessment at $Y = 1.5$m in the $XZ$-plane for all four non-uniform canyon cases. Colours as in Figure~\ref{Fig:4.10}.}
	\label{Fig:4.19}
\end{figure}

The observed changes in the flow structure in the plane $Y = 1.5$m at the pedestrian height are shown in Figure~\ref{Fig:4.18}. For the case of the step-up street canyon for $S/H = 1$, we still are in the bistable regime and the shear layers reattach closer to the front edge of the downwind building, see Figure \ref{fig:sub4.18.2}. As a consequence, small regions with ELBS number 3 appear in the centre of the canyon, Figure \ref{fig:sub4.19.2}. As we increase the street width to $S/H=2$, the shear layers move inside the canyon and reattach on the upwind face of the downwind building, see Figure \ref{fig:sub4.18.3}. This translates into a larger region of ELBS number 3 in the canyon, as shown Figure \ref{fig:sub4.19.3}. From our results we can infer that having a step-up geometry increases the chances of pedestrian discomfort with respect to a uniform canyon. On the other hand, results for the step-down geometry show qualitatively similar flow characteristics as for the uniform canyon, see Figures \ref{fig:sub4.18.4} and \ref{fig:sub4.18.3}. The pedestrian comfort charts for these configurations do not show any discernible increase or decrease in regions of ELBS number 3 or above with respect to the uniform canyon case, see Figures \ref{fig:sub4.19.4} and \ref{fig:sub4.19.5}. 

%
%

\section{Conclusion}
\label{sec:Conclusion}

We have investigated the influence of different geometric parameters defining urban street canyons on pedestrian wind comfort. The consequences of the broad features of the flow structure for  pedestrian comfort inside the street canyon mainly depend on the reattachment of the separated shear layers due to the sharp leading edges of the upstream building, and on the vortex structure in the canyon. 

We observed three general flow regimes for a variety of canyon configurations. In the stable reattachment regime, the separated shear layers reattached on the sides of the downstream building, meaning that almost all of the region inside the canyon was found to be comfortable for pedestrians. In the bistable flow regime, a part of the separated shear layers rolls up into the canyon, causing a large zone of discomfort for pedestrians. For the stable synchronized regime, where separated shear layers roll up strongly in to the canyon without reattachment, a large region of discomfort at pedestrian height was observed. 

We investigated when these three flow regimes occurred for canyons of different type. For uniform canyons, regular and medium canyons are in the stable reattachment regime which is good for pedestrian comfort. As we move towards the avenue type, we enter the bistable regime which is bad for pedestrian comfort. For all these cases, a transition from a medium to a short canyon is bad for pedestrian comfort.


Comparing non-uniform canyons to uniform canyons, there is a simple story: step-up canyons increase the size of the region of pedestrian discomfort while step-down canyons decrease the size of that region. This is an argument for town planners to encourage uniform street canyons since what is a step-down canyon one day will be a step-up canyon another when the wind changes direction. It is also an argument in favour of ``smart roof'' solutions in which building roof shapes can be altered in response to changing wind directions. The potential solution is being studied by the present authors.

Likewise, in this study, only a perpendicular wind direction and buildings with flat roofs have been considered. Change in the shape of the buildings or a change in the wind angle will cause variation in the flow structure and pedestrian comfort inside the street canyon. Further investigations are recommended to focus on these areas.

\end{document}